\documentclass[12pt,a4paper]{article}
\usepackage[utf8]{inputenc}
\usepackage{epsfig}
\usepackage{mathtools}
\usepackage{amsfonts}
\usepackage{amssymb}
\usepackage{amsmath}
\usepackage{cancel}
\usepackage{color}
\usepackage{slashed}
\usepackage{cite}
\usepackage{caption}
\usepackage{subcaption}
\usepackage{comment}
\usepackage{marginnote}
\usepackage{accents}

\usepackage{geometry} 
\usepackage{graphicx}

\usepackage{multirow}           
\usepackage{array}

\allowdisplaybreaks

\newcommand{\ep}{\varepsilon}

\def\ep  {\varepsilon}

\newcommand*\pFqskip{6mu}
\catcode`,\active
\newcommand*\pFq{\begingroup
	\catcode`\,\active
	\def ,{\mskip\pFqskip\relax}%
	\dopFq
}
\catcode`\,12
\def\dopFq#1#2#3#4#5{%
	{}_{#1}F_{#2}\biggl[\genfrac..{0pt}{}{#3}{#4};#5\biggr]%
	\endgroup
}

\newcommand*\Fkdfskip{6mu}
\catcode`,\active
\newcommand*\Fkdf{\begingroup
	\catcode`\,\active
	\def ,{\mskip\Fkdfskip\relax}%
	\doFkdf
}
\catcode`\,12
\def\doFkdf#1#2#3#4#5#6#7#8#9{%
	F^{#1}_{#2}\biggl[\genfrac..{0pt}{}{#3}{#4} \Big| \genfrac..{0pt}{}{#5}{#6} \Big| \genfrac..{0pt}{}{#7}{#8};#9\biggr]%
	\endgroup
}

\begin{document}

\begin{center}
	
\vspace{2cm}		
		
{\Large\bf Non-planar elliptic vertex} \vspace{1cm}

{\large M.A. Bezuglov$^{1,2,3}$ and A.I. Onishchenko$^{1,3,4}$}\vspace{0.5cm}
		
{\it
$^1$Bogoliubov Laboratory of Theoretical Physics,  Joint
Institute for Nuclear Research,\\ Dubna, Russia, \\
$^2$Moscow Institute of Physics and Technology (State University), Dolgoprudny, Russia, \\
$^3$Budker Institute of Nuclear Physics, Novosibirsk, Russia, \\
$^4$Skobeltsyn Institute of Nuclear Physics,  Moscow State University, Moscow, Russia}
\vspace{1cm}

\abstract{We consider the problem of obtaining higher order in regularization parameter $\epsilon$ analytical results for master integrals with elliptics. The two commonly employed methods are provided by the use of differential equations and direct integration of parametric representations in terms of iterated integrals. Taking non-planar elliptic vertex as an example we show that in addition to two mentioned methods one can use analytical solution of differential equations in terms of power series. Moreover, in the last case it is possible to obtain the exact in $\epsilon$ results expressible either in terms of generalized hypergeometric or Kampé de Fériet functions
}

\end{center}

\newpage

\section{Introduction}

The evaluation of multiloop Feynman diagrams is already a mature subject. There are already a lot of analytical methods one can choose from, such as differential equation method \cite{diffeqn1,diffeqn2,diffeqn3,diffeqn4,diffeqn5,epform1,epform2} or direct integration of parametric representations \cite{BrownModuli,linear-reducibility-1,linear-reducibility-2,PanzerAlgorithms,directint1,directint2,directint3}. Often the results for Feynman diagrams can be written in terms of a well studied class of functions known as multiple polylogarithms (MPLs) \cite{polylog1,polylog2,polylog3}. In all these cases it is known that either the corresponding differential system reduces to the so called $\varepsilon$-form \cite{epform1,epform2,epform-criterium} or corresponding parametric representation posses the property of linear reducibility \cite{linear-reducibility-1,linear-reducibility-2}. When going beyond multiple polylogarithms these methods together with a class of functions need extension. For example, in differential equation method besides a new class of functions one may either allow for non-algebraic transformation to $\varepsilon$-form \cite{epform-elliptics} or use a notion of regular basis for non-polylogarithmic integrals \cite{ep-regular-basis}. In the case of direct integration the extension of integration algorithms to a new class of functions is also required  \cite{Broedel:2017kkb,Broedel:2017siw,Broedel:2019hyg,LinearReducibledEllipticFeynmanIntegrals}. We also have  a lot of progress in understanding simplest functions beyond multiple polylogarithms, the so-called elliptic polylogarithms (EPLs) \cite{Beilinson:1994,Wildeshaus,Levin:1997,Levin:2007,Enriquez:2010,Brown:2011,Bloch:2013tra,Adams:2014vja,Bloch:2014qca,Adams:2015gva,Adams:2015ydq,Adams:2016xah,Remiddi:2017har,Broedel:2017kkb,Broedel:2017siw,Broedel:2018iwv,Broedel:2018qkq,Broedel:2019hyg,Broedel:2019tlz,Bogner:2019lfa,Broedel:2019kmn,Walden:2020odh,Weinzierl:2020fyx,kites-elliptic}.
Further extensions can include cases with several elliptic curves \cite{Adams:2018bsn,Adams:2018kez}
or one can meet completely new functions, such as in \cite{Bloch:2014qca,Primo:2017ipr,Bourjaily:2017bsb,Bourjaily:2018ycu,Bourjaily:2018yfy,kites-elliptic,Bonisch:2021yfw}.

In the present paper we will use one of the simplest elliptic Feynman diagrams, the non-planar elliptic vertex, as a laboratory to further study both differential equation method and direct integration. Specifically, we will be interested in obtaining higher order terms of the  expansion in dimensional regularization parameter $\varepsilon$  in terms of iterated integrals with algebraic kernels. Also we will see, that in addition to two mentioned methods one can use Frobenius method\footnote{For previous applications of Frobenius method in the context of Feynman diagrams see for example \cite{Frobenius1,Frobenius2,Frobenius3,Frobenius4,Frobenius5,KKOVelliptic2,Bonisch:2021yfw}} to find analytical power series solution of differential equation system. Moreover, in the last case it is possible to obtain solution exact in $\varepsilon$.

We organized paper as follows. In the next section we consider both differential equation and direct integration methods. The obtained results are expressed through iterated integrals with algebraic kernels \cite{kites-elliptic,banana-elliptic} and are valid up to $\mathcal{O} (\varepsilon^2)$ order. Next, in section \ref{Frobenius} we present the details of obtaining exact in $\varepsilon$  power series solution with Frobenius method. Finally, in section \ref{conclusion&Discussion} we put a comparative analysis of three methods used. Appendices contain notation and extra details.

\section{Differential equations and direct integration}

\begin{figure}[h]
	\center{\includegraphics[width=0.7\textwidth]{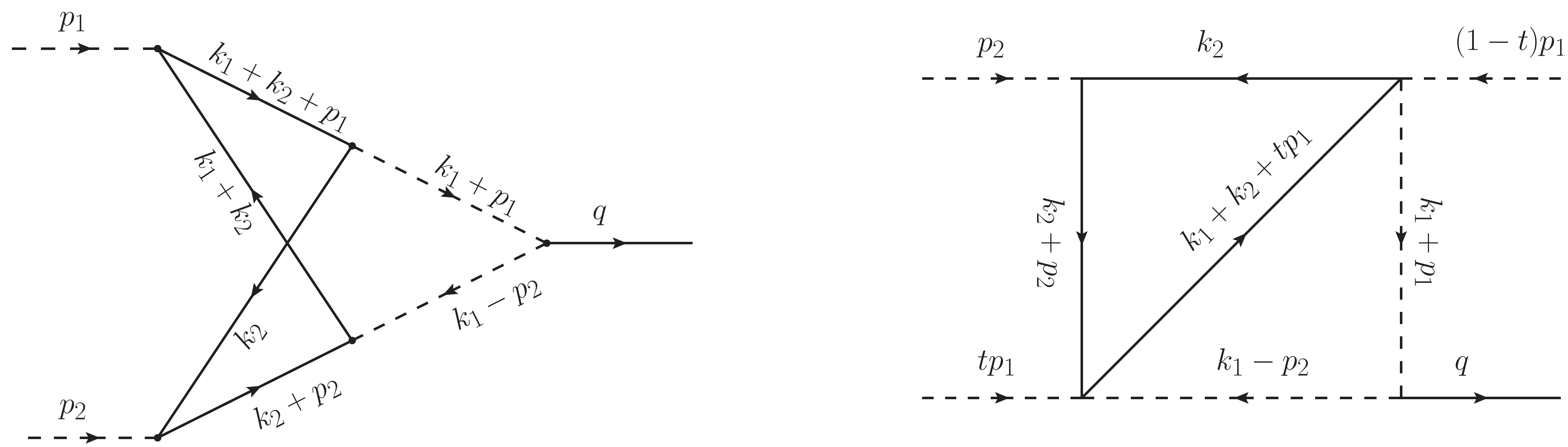}}
	\caption{Non-planar elliptic vertex on the left and effective diagram obtained by combining two propagators with Feynman parameter on the right. }
	\label{figs:nonplanar-vertex}
\end{figure}

There are basically two common methods employed for analytical solution of elliptic master integrals. The latter are given by differential equations \cite{diffeqn1,diffeqn2,diffeqn3,diffeqn4,diffeqn5} and direct integration of parametric representations\footnote{In principle one can also use Mellin-Barnes representation.} in terms of iterated integrals \cite{BrownModuli,linear-reducibility-1,linear-reducibility-2,PanzerAlgorithms,directint1,directint2,directint3,BognerMPL,PolyLogTools,Broedel:2017kkb,Broedel:2017siw,Broedel:2019hyg}. Both of these methods can be used to obtain results up to in principle arbitrary order in the parameter of dimensional regularization $\varepsilon$ ($d=4-2\varepsilon$). To illustrate their use let us consider non-planar elliptic vertex\footnote{Note, that for applications only the leading $\varepsilon$  expressions for two top elliptic master integrals are sufficient.} shown in Fig. \ref{figs:nonplanar-vertex} \footnote{A similar vertex diagram but with a different mass distribution for propagators was considered in \cite{Aglietti:2007as}.}. The master integrals for this vertex diagram can be conveniently described in terms of the following integral family:
\begin{equation}
I^{v}_{a_1,...,a_6}(s)=e^{2\varepsilon\gamma_E}\int\frac{d^dk_1d^dk_2}{(i\pi^{d/2})^2}\frac{1}{D_1^{a_1}D_2^{a_2}D_3^{a_3}D_4^{a_4}D_5^{a_5}D_6^{a_6}}\, , \label{eqns:EV}
\end{equation}
where
\begin{multline}
D_1 =1-(k_1+k_2+p_1)^2,  D_2 = 1- (k_1+k_2), D_3 = 1-k_2^2, \\  D_4 = 1-(k_2+p_2)^2,  D_5 = -(k_1+p_1)^2,  D_6 =-(k_1-p_2)^2 ,
\label{eqns:EVProp} 
\end{multline}
and  kinematics is given by $p_1^2=0,~p_2^2=0$ and $q^2=4s$. 

This family contains two elliptic master integrals which can be chosen as $I^{v}_{1,1,1,1,1,1}$ and  $I^{v}_{2,1,1,1,1,1}$. The rest of master integrals are non-elliptic and can be conveniently expressed in terms of multiple polylogarithms.

\subsection{Differential equations for effective master integrals}

The use of differential equations to evaluate these master integrals was already considered in \cite{nonplanar-vertex}. In the latter paper authors used Euler's variation of constants to solve second order inhomogeneous differential equations for the mentioned elliptic master integrals. The results obtained in \cite{nonplanar-vertex} are valid to leading order in $\varepsilon$, but can be also extended to higher orders.  However, the results obtained in this way will contain multiple integrals with complex elliptic kernels. To have a higher order $\varepsilon$ solution in terms of iterated integrals with simpler algebraic kernels we will follow the method described in \cite{kites-elliptic,banana-elliptic}. First, we combine two propagators from the set \eqref{eqns:EVProp} using Feynman parameter trick as introduced in\footnote{For earlier application of such procedure in non-elliptic context see \cite{effectivemass1,effectivemass2,effectivemass3,effectivemass4,effectivemass5} and references therein.} \cite{KKOVelliptic1,KKOVelliptic2,LinearReducibledEllipticFeynmanIntegrals}: 
\begin{equation}
\frac{1}{A^mB^n}=\frac{\Gamma(n+m)}{\Gamma(m)\Gamma(n)}\int\limits_0^1\frac{t^{m-1}(1-t)^{n-1}dt}{(tA+(1-t)B)^{n+m}},
\end{equation}
When this trick is done for a sunset integral with equal masses the choice of two particular propagators is not important. This is obviously not true for an elliptic vertex. Empirically we have found that the best choice in this case is  $A=D_1$ and $B=D_2$. Then the expressions for two elliptic master integrals are
\begin{equation}
I^{v}_{1,1,1,1,1,1}(s)=e^{2\varepsilon\gamma_E}\int\limits_0^1 dt \int\frac{d^dk_1d^dk_2}{(i\pi^{d/2})^2}\frac{1}{D_{12}^2D_3D_4D_5D_6},
\label{master1}
\end{equation}
\begin{equation}
I^{v}_{2,1,1,1,1,1}(s)=2e^{2\varepsilon\gamma_E}\int\limits_0^1 dt(1-t) \int\frac{d^dk_1d^dk_2}{(i\pi^{d/2})^2}\frac{1}{D_{12}^3D_3D_4D_5D_6},
\label{master2}
\end{equation}
where
\begin{equation}
D_{12}= 1-(k_1+k_2)^2-2t(k_1+k_2)p_1.
\end{equation}
Next, the expressions for two loop integrals in $q_1$ and $q_2$ can then be conveniently obtained with the use of differential equation method. To do this we define the family of effective two loop integrals as (see right part of Fig. \ref{eqns:EV}):
\begin{equation}
I^{\rm eff}_{b_{12},b_3,...,b_6}(s,t)=e^{2\varepsilon\gamma_E}\int\frac{d^dk_1d^dk_2}{(i\pi^{d/2})^2}\frac{1}{D_{12}^{b_{12}}D_3^{b_3}D_4^{b_4}D_5^{b_5}D_6^{b_6}}
\label{eqns:effFamily}
\end{equation}

The vector of thirteen IBP master integrals for this family obtained as a result of IBP reduction \cite{IBP1,IBP2} can be chosen in the following form:
\begin{align}
{\bf I}_{\rm IBP}^{\rm eff} = \{&I^{\rm eff}_{1,0,1,0,0},& &I^{\rm eff}_{0,0,1,1,1},& &I^{\rm eff}_{1,0,1,1,0},& &I^{\rm eff}_{2,0,1,1,0},& &I^{\rm eff}_{1,1,0,0,1},& &I^{\rm eff}_{2,1,0,0,1},& &I^{\rm eff}_{1,0,1,1,1}, \nonumber \\ &I^{\rm eff}_{2,0,1,1,1},& &I^{\rm eff}_{1,1,0,1,1},& &I^{\rm eff}_{2,1,0,1,1},&   &I^{\rm eff}_{1,1,1,0,1},& &I^{\rm eff}_{1,1,1,1,0},& &I^{\rm eff}_{2,1,1,1,1}\}^{\top}
\label{eqns:effIBP}
\end{align}
This set of master integrals is not complete. The complete basis of master integrals for integral family \eqref{eqns:effFamily} should include two more masters, which for example can be chosen as $I^{\rm eff}_{1,1,1,1,1}$ and $I^{\rm eff}_{1,2,1,1,1}$. However, in order to find expressions for  master integrals \eqref{master1} and \eqref{master2} we only need to know  expressions for $I^{\rm eff}_{2,1,1,1,1}$ and $I^{\rm eff}_{3,1,1,1,1}$. It turns out that 
neither of these integrals nor differential system for integrals in Eq. \eqref{eqns:effIBP} depend on  $I^{\rm eff}_{1,1,1,1,1}$ or $I^{\rm eff}_{1,2,1,1,1}$. Note also, that the complete basis of master integrals for integral family \eqref{eqns:effFamily} is not interesting by itself and thus we can safely exclude those two master integrals from later consideration. It is a big simplification because despite the fact that the results for "unnecessary" master integrals can be expressed through the same functions as  masters from Eq. \eqref{eqns:effIBP} the corresponding differential system is more cumbersome. The differential system for a set of master integrals \eqref{eqns:effIBP} overs $s$ is then obtained with the use of IBP relations. Moreover, this differential system can be further reduced to  $\varepsilon$-form \cite{epform1,epform2} through the transformation to the basis of canonical master integrals\footnote{The derivation of IBP identities and subsequent reduction of system of differential equations to $\varepsilon$-form were performed with the use of  \texttt{LiteRed} \cite{LiteRed1,LiteRed2}  and \texttt{Libra} \cite{Libra} packages.} $\widetilde{I}^{\rm eff}_{i}(s,t)$ $(i=1,\ldots 13)$: 
\begin{equation}
\label{eqns:eqGeneral}
\frac{d \widetilde{{\bf I}}^{\rm eff}}{ds}=\varepsilon({\bf M}+{\bf W})\cdot \widetilde{{\bf I}}^{\rm eff}\, , \qquad {\bf I}_{\rm IBP}^{\rm eff} = {\bf T}\cdot \widetilde{{\bf I}}^{\rm eff}
\end{equation}
where ($\bar{t}=1-t$)
\begin{multline}
{\bf M} =\frac{\mathcal{M}^{z_1}_{1/\bar{t}}z_1}{s-1/\bar{t}}+\frac{\mathcal{M}^{z_2}_{1/t}z_2}{s-1/t}+\frac{\mathcal{M}^{z_3}_{-1/t}z_3}{s+1/t}+\frac{\mathcal{M}^{z_4}_{-1/\bar{t}}z_4}{s+1/\bar{t}}+\frac{\mathcal{M}^{z_5}_{-1/\bar{t}t}z_5}{s+1/\bar{t}t}+\frac{\mathcal{M}^{1}_{-1/t}}{s+1/t}+\frac{\mathcal{M}^{1}_{1/t}}{s-1/t}+
\\
+\frac{\mathcal{M}_{1/\bar{t}}^{1}}{s-1/\bar{t}}+
\frac{\mathcal{M}_{-1/\bar{t}}^{1}}{s+1/\bar{t}}+\frac{\mathcal{M}_{-1/\bar{t}t}^{1}}{s+1/\bar{t}t}+\frac{\mathcal{M}_{0}^{1}}{s},
\end{multline}
\begin{multline}
{\bf W}=\frac{\mathcal{W}^{z_4z_5}_{-1/\bar{t}t}z_4z_5}{s+1/\bar{t}t}+\frac{\mathcal{W}^{z_4z_5}_{-1/\bar{t}}z_4z_5}{s+1/\bar{t}}+\frac{\mathcal{W}^{z_3z_5}_{-1/\bar{t}t}z_3z_5}{s+1/\bar{t}t}+\frac{\mathcal{W}^{z_3z_5}_{-1/t}z_3z_5}{s+1/t}+\frac{\mathcal{W}^{z_1z_3}_{-1/t}z_1z_3}{s+1/t}+\\+\frac{\mathcal{W}^{z_1z_3}_{1/\bar{t}}z_1z_3}{s-1/\bar{t}}+\frac{\mathcal{W}^{z_2z_4}_{1/t}z_2z_4}{s-1/t}+\frac{\mathcal{W}^{z_2z_4}_{-1/\bar{t}}z_2z_4}{s+1/\bar{t}},
\end{multline}
and
\begin{equation}
\label{eqns:roots}
z_1=\sqrt{\frac{s-1/\bar{t}}{s}}, ~ z_2=\sqrt{\frac{s-1/t}{s}}, ~ z_3=\sqrt{\frac{s+1/t}{s}}, ~ z_4=\sqrt{\frac{s+1/\bar{t}}{s}}, ~ z_5=\sqrt{\frac{s+1/\bar{t}t}{s}}.
\end{equation}
Here,  $\mathcal{M}_a^b$ and $\mathcal{W}_a^b$ are $13 \times 13$ matrix coefficients depending on $t$. For example
\begin{equation}
\mathcal{M}_{0}^{1} = \left(
\begin{array}{ccccccccccccc}
0 & 0 & 0 & 0 & 0 & 0 & 0 & 0 & 0 & 0 & 0 & 0 & 0 \\
0 & -1 & 0 & 0 & 0 & 0 & 0 & 0 & 0 & 0 & 0 & 0 & 0 \\
7-4 t & 0 & -1 & 14-8 t & 0 & 0 & 0 & 0 & 0 & 0 & 0 & 0 & 0 \\
1 & 0 & 0 & 1 & 0 & 0 & 0 & 0 & 0 & 0 & 0 & 0 & 0 \\
4 t+3 & 0 & 0 & 0 & -1 & 8 t+6 & 0 & 0 & 0 & 0 & 0 & 0 & 0 \\
1 & 0 & 0 & 0 & 0 & 1 & 0 & 0 & 0 & 0 & 0 & 0 & 0 \\
0 & 0 & 0 & 0 & 0 & 0 & -1 & 2-4 t & 0 & 0 & 0 & 0 & 0 \\
0 & 0 & 0 & 0 & 0 & 0 & 0 & 1 & 0 & 0 & 0 & 0 & 0 \\
0 & 0 & 0 & 0 & 0 & 0 & 0 & 0 & -1 & 4 t-2 & 0 & 0 & 0 \\
0 & 0 & 0 & 0 & 0 & 0 & 0 & 0 & 0 & 1 & 0 & 0 & 0 \\
0 & 0 & 0 & 0 & 0 & 0 & 0 & 0 & 0 & 0 & 0 & 0 & 0 \\
0 & 0 & 0 & 0 & 0 & 0 & 0 & 0 & 0 & 0 & 0 & 0 & 0 \\
0 & 0 & 0 & 0 & 0 & 0 & 0 & 0 & 0 & 0 & 0 & 0 & -1 \\
\end{array}
\right)
\end{equation}
and other coefficient matrices together with the transformation matrix $\bf T$ to the canonical basis can be found in accompanying {\it Mathematica} notebook. 

Having reduced the differential system to $\varepsilon$-form its solution can be straightforwardly written in terms of iterated integrals with algebraic kernels. The boundary conditions for the differential system are obtained from the requirement, that the
expansion of master integrals \eqref{eqns:effIBP} in $s$ do not contain terms with non-integer powers of $s$. Also, we require that some master integrals contain only logarithmic singularities at $s=0$. Another way to obtain these boundary conditions is to solve the system of differential equations with respect to the parameter $t$ at $s=0$ as was done in \cite{kites-elliptic}. The expressions for effective master integrals obtained in this way can be found in Appendix \ref{effective-masters}. Next, using the notation from Appendix \ref{notation} and expressions for corresponding effective master integrals the results for two elliptic master integrals \eqref{master1} and \eqref{master2} can be written as
\begin{multline}
I^{v}_{1,1,1,1,1,1}(s) =\frac{1}{8s}\Bigg[ J\left(N_{-1,0}^{z_5},\omega _{-1/\bar{t}t}^{z_3 z_5},\omega
_{-1/t}^{z_3},\omega _0;s\right)-J\left(N_{-1,1}^{z_5},\omega _{-1/\bar{t}t}^{z_3 z_5},\omega _{-1/t}^{z_1 z_3},\omega
_{1/\bar{t}}^{z_1};s\right)
\\
+J\left(N_{-1,2}^{z_5},\omega
_{-1/t}^{z_3 z_5},\omega _{-1/t}^{z_1 z_3},\omega
_{1/\bar{t}}^{z_1};s\right)+J\left(N_{0,-1}^{z_5},\omega _{-1/\bar{t}t}^{z_4 z_5},\omega _{-1/\bar{t}}^{z_4},\omega
_0;s\right)
\\
-J\left(N_{0,0}^{z_5},\omega _{-1/\bar{t}t}^{z_5},\omega
_{1/t}^{z_2},\omega _{1/t}^{z_2};s\right)-J\left(N_{0,0}^{z_5},\omega
_{-1/\bar{t}t}^{z_5},\omega _{1/\bar{t}}^{z_1},\omega
_{1/\bar{t}}^{z_1};s\right)
\\
-J\left(N_{0,0}^{z_5},\omega _{-1/\bar{t}t}^{z_3 z_5},\omega _{1/\bar{t}}^{z_1 z_3},\omega
_{1/\bar{t}}^{z_1};s\right)-J\left(N_{0,0}^{z_5},\omega _{-1/\bar{t}t}^{z_4 z_5},\omega _{1/t}^{z_2 z_4},\omega
_{1/t}^{z_2};s\right)
\\
+J\left(N_{0,1}^{z_5},\omega _{-1/t}^{z_3
	z_5},\omega _{1/\bar{t}}^{z_1 z_3},\omega
_{1/\bar{t}}^{z_1};s\right)-J\left(N_{1,-1}^{z_5},\omega _{-1/\bar{t}t}^{z_4 z_5},\omega _{-1/\bar{t}}^{z_2 z_4},\omega
_{1/t}^{z_2};s\right)
\\
+J\left(N_{1,0}^{z_5},\omega
_{-1/\bar{t}}^{z_4 z_5},\omega _{1/t}^{z_2 z_4},\omega
_{1/t}^{z_2};s\right)
+J\left(N_{2,-1}^{z_5},\omega
_{-1/\bar{t}}^{z_4 z_5},\omega _{-1/\bar{t}}^{z_2 z_4},\omega
_{1/t}^{z_2};s\right)
\\
-
J\left(N_{-1,1}^{z_5},\omega
_{-1/t}^{z_3 z_5},\omega _{-1/t}^{z_3},\omega _0;s\right)-J\left(N_{1,-1}^{z_5},\omega
_{-1/\bar{t}}^{z_4 z_5},\omega _{-1/\bar{t}}^{z_4},\omega
_0;s\right)
\\
+
(\log (4)-i \pi )\Big\{ J\left(N_{-1,0}^{z_5},\omega
_{-1/\bar{t}t}^{z_3 z_5},\omega _{-1/t}^{z_3};s\right)- J\left(N_{1,-1}^{z_5},\omega
_{-1/\bar{t}}^{z_4 z_5},\omega _{-1/\bar{t}}^{z_4};s\right)
\\
-J\left(N_{-1,1}^{z_5},\omega _{-1/t}^{z_3 z_5},\omega
_{-1/t}^{z_3};s\right)+J\left(N_{0,-1}^{z_5},\omega _{-1/\bar{t}t}^{z_4 z_5},\omega
_{-1/\bar{t}}^{z_4};s\right) \Big\}\Bigg] + \mathcal{O} (\varepsilon)
\end{multline}
and
\begin{multline}
I^{v}_{2,1,1,1,1,1}(s) =\frac{1}{64s}\Bigg[-2 s J\left(K_{0,1}^{z_4},\omega
_{-1/\bar{t}}^{z_4},\omega _0;s\right)+2 s J\left(K_{0,2}^{z_4},\omega
_{1/t}^{z_2 z_4},\omega _{1/t}^{z_2};s\right)
\\
+2 s
J\left(K_{1,1}^{z_4},\omega _{-1/\bar{t}}^{z_2 z_4},\omega
_{1/t}^{z_2};s\right)-2 J\left(N_{0,0}^{\text{}},\omega
_{1/t}^{z_2},\omega _{1/t}^{z_2};s\right)
\\
-2
J\left(N_{0,0},\omega _{1/\bar{t}}^{z_1},\omega
_{1/\bar{t}}^{z_1};s\right)+2 s J\left(\Omega _{1,0}^{z_3},\omega
_{-1/t}^{z_3},\omega _0;s\right)
\\
-2 s J\left(\Omega _{1,1}^{z_3},\omega
_{-1/t}^{z_1 z_3},\omega _{1/\bar{t}}^{z_1};s\right)-2 s
J\left(\Omega _{2,0}^{z_3},\omega _{1/\bar{t}}^{z_1 z_3},\omega
_{1/\bar{t}}^{z_1};s\right)
\\
+J\left(F_{-1,0}^{z_5},\omega
_{-1/t}^{z_3 z_5},\omega _{-1/t}^{z_3},\omega
_0;s\right)-J\left(F_{-1,0}^{z_5},\omega _{-1/\bar{t}t}^{z_3 z_5},\omega
_{-1/t}^{z_3},\omega _0;s\right)
\\
+J\left(F_{-1,0}^{z_5},\omega _{-1/\bar{t}t}^{z_3 z_5},\omega _{-1/t}^{z_1 z_3},\omega
_{1/\bar{t}}^{z_1};s\right)-J\left(F_{-1,2}^{z_5},\omega
_{-1/t}^{z_3 z_5},\omega _{-1/t}^{z_1 z_3},\omega
_{1/\bar{t}}^{z_1};s\right)
\\
-J\left(F_{0,0}^{z_5},\omega _{-1/t}^{z_3 z_5},\omega
_{-1/t}^{z_3},\omega _0;s\right)+J\left(F_{0,0}^{z_5},\omega _{-1/\bar{t}t}^{z_5},\omega _{1/t}^{z_2},\omega
_{1/t}^{z_2};s\right)
\\
+J\left(F_{0,0}^{z_5},\omega _{-1/\bar{t}t}^{z_5},\omega _{1/\bar{t}}^{z_1},\omega
_{1/\bar{t}}^{z_1};s\right)-J\left(F_{0,0}^{z_5},\omega _{-1/\bar{t}t}^{z_3 z_5},\omega _{-1/t}^{z_1 z_3},\omega
_{1/\bar{t}}^{z_1};s\right)
\\
+J\left(F_{0,0}^{z_5},\omega _{-1/\bar{t}t}^{z_3 z_5},\omega _{1/\bar{t}}^{z_1 z_3},\omega
_{1/\bar{t}}^{z_1};s\right)+J\left(F_{0,0}^{z_5},\omega _{-1/\bar{t}t}^{z_4 z_5},\omega _{1/t}^{z_2 z_4},\omega
_{1/t}^{z_2};s\right)
\\
-J\left(F_{0,-1}^{z_5},\omega _{-1/\bar{t}t}^{z_4 z_5},\omega _{-1/\bar{t}}^{z_4},\omega
_0;s\right)-J\left(F_{0,1}^{z_5},\omega _{-1/t}^{z_3
	z_5},\omega _{1/\bar{t}}^{z_1 z_3},\omega
_{1/\bar{t}}^{z_1};s\right)
\\
+J\left(F_{1,-1}^{z_5},\omega
_{-1/\bar{t}}^{z_4 z_5},\omega _{-1/\bar{t}}^{z_4},\omega
_0;s\right)+J\left(F_{1,-1}^{z_5},\omega _{-1/\bar{t}t}^{z_4 z_5},\omega
_{-1/\bar{t}}^{z_2 z_4},\omega
_{1/t}^{z_2};s\right)
\\
-J\left(F_{1,0}^{z_5},\omega
_{-1/\bar{t}}^{z_4 z_5},\omega _{1/t}^{z_2 z_4},\omega
_{1/t}^{z_2};s\right)-J\left(F_{2,-1}^{z_5},\omega
_{-1/\bar{t}}^{z_4 z_5},\omega _{-1/\bar{t}}^{z_2 z_4},\omega
_{1/t}^{z_2};s\right)
\\
+
(\log (4)-i \pi )\Big\{-J\left(F_{0,-1}^{z_5},\omega
_{-1/\bar{t}t}^{z_4 z_5},\omega _{-1/\bar{t}}^{z_4};s\right) - J\left(F_{-1,0}^{z_5},\omega _{-1/\bar{t}t}^{z_3
	z_5},\omega _{-1/t}^{z_3};s\right) 
\\
+ J\left(F_{-1,1}^{z_5},\omega _{-1/t}^{z_3 z_5},\omega
_{-1/t}^{z_3};s\right) 
+ J\left(F_{1,-1}^{z_5},\omega
_{-1/\bar{t}}^{z_4 z_5},\omega _{-1/\bar{t}}^{z_4};s\right)
\\
-2 s J\left(K_{0,1}^{z_4},\omega
_{-1/\bar{t}}^{z_4};s\right)+2 s J\left(\Omega _{1,0}^{z_3},\omega
_{-1/t}^{z_3};s\right)
 \Big\}\Bigg] + \mathcal{O}(\varepsilon)
\end{multline}
In the accompanying {\it Mathematica} notebook we have also  put expressions for $\varepsilon^1$ terms in $\varepsilon$-expansion, but in principle with the presented technique we can have as many terms in $\varepsilon$-expansion as required. Note also, that the first entries in $J$-integrals are different from the remaining ones (which are all represented by $\omega$ 1-forms) due to integral representation of elliptic master integrals in terms of effective masters, see Appendix \ref{notation}.

\subsection{Integration of parametric representation}

\begin{figure}[t]
	\center{\includegraphics[width=0.7\textwidth]{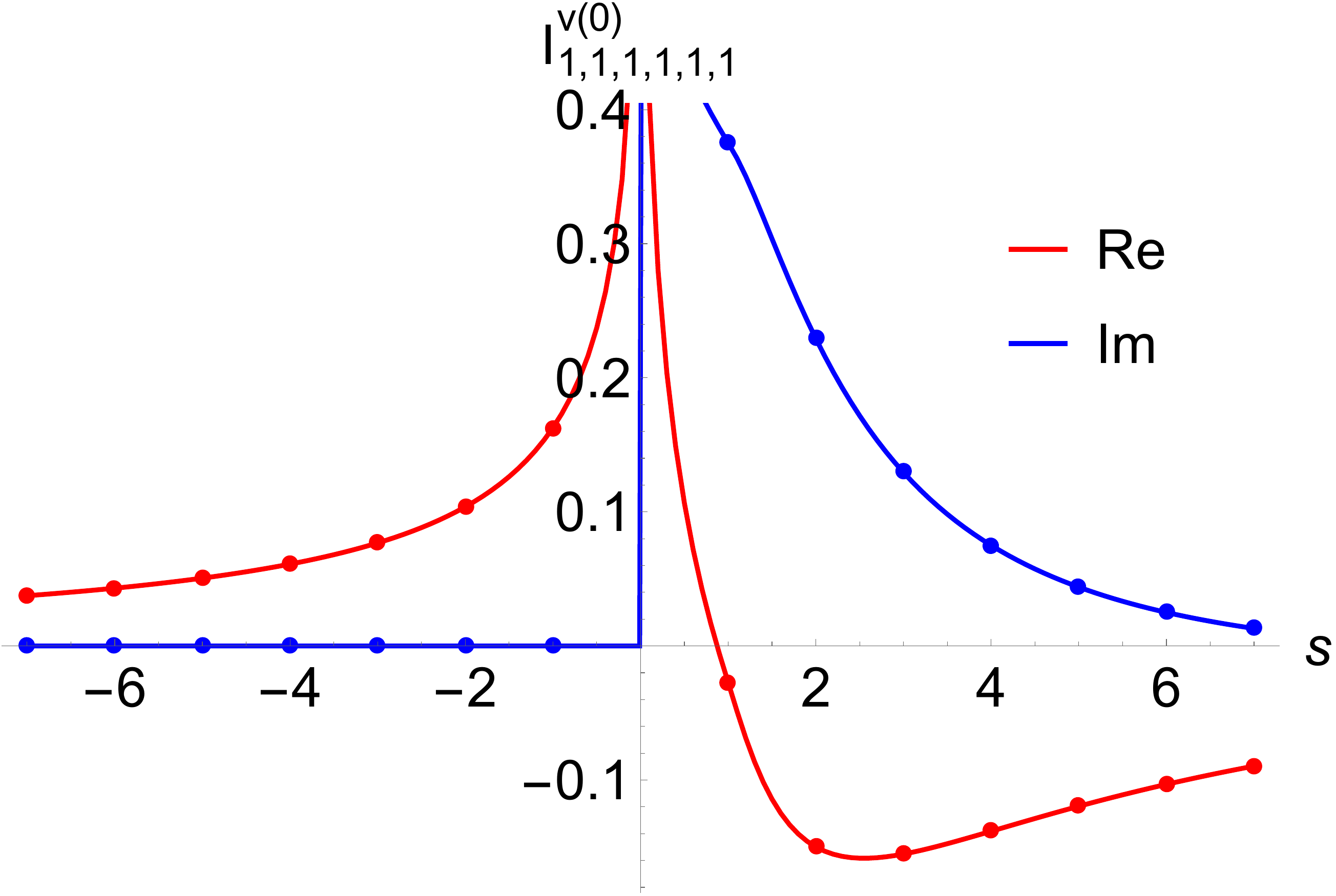}}
	\caption{Plot of the $\varepsilon^0$ correction to the $I^{v}_{1,1,1,1,1,1}$.  The solid points represent values computed numerically with the FIESTA package \cite{Fiesta4}.}
	\label{figs:ep0M1}
\end{figure}

Let us now turn to direct integration method. Such integration at leading order in $\varepsilon$ for the first elliptic master integral $I^{v}_{1,1,1,1,1,1}$ was already considered in \cite{LinearReducibledEllipticFeynmanIntegrals,Broedel:2019hyg,Broedel:2018qkq}. Here we just extend those results for the second master and next term in $\varepsilon$-expansion. The parametric representation for the first master integral $I^{v}_{1,1,1,1,1,1}$ is given by ($d=4-2\varepsilon$): 
\begin{equation}
\label{eqns:J1ParametricRep}
I^{v}_{1,1,1,1,1,1}=e^{2\gamma_E\varepsilon}\Gamma(6-d)\int_{\Delta}\left(\prod\limits_{j=1}^6 dx_j\right)\frac{U^{6-\frac{3 d}{2}}}{F^{6-d}}
\end{equation}
Here $\Delta$ is the integration domain defined as $\Delta \equiv \left\{\vec{x}~|~x_j>0,~ \sum_{j=1}^6x_j=1 \right\}$, $x_j$ are Feynman parameters and $U$ and $F$ are the first and second Symanzik polynomials. The latter are given by
\begin{equation}
U=x_4 x_3+x_5 x_3+x_4 x_5+\left(x_4+x_3\right) x_6+x_2 \left(x_4+x_5+x_6\right)+x_1
\left(x_2+x_3+x_5+x_6\right)
\end{equation}
and
\begin{multline}
F = x_1 \left(x_3 \left(-4 s x_5+x_3+2 x_5\right)+2 x_6 \left(x_3-2 s x_5\right)+x_2^2+2
\left(x_4+x_3+x_5+x_6\right) x_2\right. \\ \left. +2 x_4 \left(x_3+x_5+x_6\right)\right)
+x_6 \left(2
x_2 \left(-2 s x_4-2 s x_5+x_4+x_3\right)+\left(x_4+x_3\right) \left(-4 s
x_5+x_4+x_3\right)\right. \\ \left.+x_2^2\right) +\left(x_2+x_3+x_5+x_6\right)
x_1^2
+\left(x_2+x_4+x_3\right) \left(x_4
\left(x_2+x_3\right)+\left(x_2+x_4+x_3\right) x_5\right).
\end{multline}
The integral $I^{v}_{1,1,1,1,1,1}$ in the limit $\varepsilon\to 0$ is finite and its $\varepsilon$-expansion has the form 
\begin{equation}
I^{v}_{1,1,1,1,1,1} = \sum\limits_{i=0}^{\infty} I^{v (i)}_{1,1,1,1,1,1}\varepsilon^i.
\end{equation}
The direct integration over Feynman parameters can be greatly simplified with the use of Chen-Wu theorem \cite{ChengWu}. The latter allow us to change the integration domain $\Delta$
to any other domain $\Delta'\equiv \left\{\vec{x}~|~x_k>0,~ \sum_{I}x_k=1 \right\}$, where the summation goes over some arbitrary subset $I$ of  $\{x_1,...x_6\}$. The good choice is \cite{LinearReducibledEllipticFeynmanIntegrals,Broedel:2019hyg}: $\Delta' \equiv \left\{\vec{x}~|~x_k>0,~ \sum_{k=1}^4x_k=1 \right\} $.  Changing in Eq. \eqref{eqns:J1ParametricRep} $\Delta \to \Delta'$ and substituting $x_1 = 1-x_2-x_3-x_4$ to leading order in $\varepsilon$ we find
\begin{equation}
I_{1,1,1,1,1,1}^{v(0)}=2\int\limits_0^{1}dx_2\int\limits_0^{1-x_2}dx_3\int\limits_0^{1-x_2-x_3}dx_4\int\limits_0^{\infty}dx_5\int\limits_0^{\infty}dx_6 \left. \frac{1}{F^2}\right|_{x_1 = 1-x_2-x_4-x_3}.
\end{equation}
The first two integrations over the variables $x_6$ and $x_5$ are trivial, such that
\begin{multline}
I_{1,1,1,1,1,1}^{v(0)}=2\int\limits_0^{1}dx_2\int\limits_0^{1-x_2}dx_3\int\limits_0^{1-x_2-x_3}dx_4\frac{1}{(4 s x_4 x_3+1) (4 s x_2 (x_2+x_4+x_3-1)-1)}\\
\times\Bigg[G\left(0;\frac{(x_2+x_3-1) (x_2+x_3)}{4 s x_3
	(x_2+x_3-1)+1}\right) -G\left(\frac{1}{4 s x_2};x_4\right)\\ -G\left(\frac{-4 s
	x_3^2+4 s x_3-4 s x_2 x_3-1}{4 s x_3};x_4\right)-G\left(0;\frac{1}{4 s}\right)\Bigg]\, ,
\end{multline}
where we have used the fact that all multiple polylogarithms (MPLs) of weight 1 are just ordinary logarithms $G(a;b)=\log\left(1-\frac{b}{a}\right),~ a\ne 0$ and $G(0;b)=\log(b)$. The next integration over $x_4$ is also easy and goes with the application of MPLs definition. After this step we get: 
\begin{multline}
\label{eqns:J1afterx5x6Ints}
I_{1,1,1,1,1,1}^{v(0)}=2\int\limits_0^{1}dx_2\int\limits_0^{1-x_2}dx_3\frac{1}{4 s \left(4 s x_3 x_{23} x_2+x_2+x_3\right)}\\ \times\Bigg[
G\left(-\frac{1}{4 s x_3},\frac{1}{4 s
	x_2};x_{23}\right)+G\left(-\frac{1}{4 s x_3},x_{23}-\frac{1}{4 s
	x_3};x_{23}\right)
\\   
-\left(G\left(0;\frac{1}{4 s}\right)-G\left(0;-\frac{\left(x_2+x_3\right)
	x_{23}}{1-4 s x_3 x_{23}}\right)\right)\left(G\left(\frac{1}{4 s
	x_2}+x_{23};x_{23}\right)-G\left(-\frac{1}{4 s
	x_3};x_{23}\right)\right)
\\
-G\left(\frac{1}{4 s x_2}+x_{23},\frac{1}{4 s
	x_2};x_{23}\right)-G\left(\frac{1}{4 s x_2}+x_{23},x_{23}-\frac{1}{4 s
	x_3};x_{23}\right)\Bigg]\, ,   
\end{multline}
where $x_{23} = 1-x_2-x_3$. To integrate this expression over $x_3$ with the use of MPLs definition we will need to first rewrite all MPLs with $x_3$ present either in indexes or in argument in terms of MPLs $G(a_1,...,a_n; x_3)$ from argument $x_3$ with all indexes $a_i$ independent from $x_3$. In what follows, we will call such a form canonical. The reduction to this form can be achieved either thought the application of symbols and coproducts  \cite{symbolsapp1,symbolsapp2,symbolsapp3} or by recursively\footnote{The recursion stops as at each iteration step the weight of MPL is reduced by one.} differentiating the required MPLs with respect to $x_3$ and integrating them back. In both cases there is a problem of determing the constants of integration. The latter could be determined either numerically with PSLQ as in \cite{directint3} or analytically as in \cite{PanzerAlgorithms}. Making the variable change \footnote{This replacement slightly simplifies the subsequent calculation \cite{Broedel:2019hyg}} $x_2 = t - x_3$ and rewriting all MPLs in the canonical form for $x_3$ we find:   
\begin{multline}
I_{1,1,1,1,1,1}^{v(0)}=2\int\limits_0^{1}dt\int\limits_0^{t}dx_3\frac{1}{4 s \left(-4 s \bar{t} x_3^2+\frac{4 s x_3}{y^2}+t\right)}\\ \times\Bigg[
G\left(0;-\frac{1}{y^2}\right) \left(-\frac{1}{2}
G\left(\frac{1}{4 s \bar{t}}+t;x_3\right)-\frac{1}{2} G\left(\frac{1}{4 s
	\bar{t}};x_3\right)\right) +\frac{1}{2} G\left(\frac{y^2}{4 s};1\right) G\left(\frac{1}{4 s
	\bar{t}};x_3\right) \\ +G\left(0,\frac{1}{4 s}\right) \left(\frac{1}{2}
G\left(\frac{1}{4 s \bar{t}}+t;x_3\right)+\frac{1}{2} G\left(\frac{1}{4 s
	\bar{t}};x_3\right)-\frac{1}{2} G\left(\frac{y^2}{4
	s}+1;1\right)\right)
\\
+G\left(0,\frac{1}{4 s \bar{t}};x_3\right)+G\left(t,\frac{1}{4 s
	\bar{t}}+t;x_3\right)+\frac{1}{2} G\left(\frac{1}{4 s \bar{t}}+t,\frac{1}{4 s
	\bar{t}};x_3\right)+\frac{1}{2} G\left(\frac{1}{4 s \bar{t}},\frac{1}{4 s
	\bar{t}}+t;x_3\right)
\\
+G(t,x_3) G\left(\frac{y^2}{4 s};1\right)+\frac{1}{2}
G\left(0,-\frac{1}{y^2}\right) G\left(\frac{y^2}{4 s}+1;1\right)-\frac{1}{2}
G\left(\frac{y^2}{4 s}+1,\frac{y^2}{4 s};1\right)\Bigg].
\end{multline}
where $\bar{x}_3 = t - x_3$, $y^2 = 1/t\bar{t}$ and $\bar{t} =1-t$. Now this expression can be directly integrated over the variable $x_3$ and we get 
\begin{multline}
\label{eqns:J1almostFinal}
I_{1,1,1,1,1,1}^{v(0)}=2\int\limits_0^{1}dt\frac{z^2(1-z)^2}{y^2(2z-1)}\Bigg[\Big(
G(z;1) G\left(z^2-z+1;1\right) +G\left(1-z,z^2-z+1;1\right)
\\ 
-G\left(z,z^2-z+1;1\right)+G(1-z,(1-z
) z;1) -G(z,(1-z) z;1)\Big) G\left(0;\frac{(z-1)
	z}{y^2}\right)
	\\ 
	+G\left(0;-\frac{1}{y^2}\right) \Big(G\left(z^2-z+1;1\right)\left(G(1-z;1)-G(z;1)\right)-G\left(1-z,z^2-z+1;1\right)
\\
 +G\left(z,z^2-z+1;1\right)-G(1-z,(1-z) z;1)+G(z,(1-z) z;1)\Big)\\
+G(1-z;1) \left(G\left(z^2-z+1;1\right)
\left(-G\left(0;\frac{(z-1) z}{y^2}\right)\right)-G\left(z^2-z+1,(z-1)
z;1\right)\right)\\ +G(z;1) G\left(z^2-z+1,(z-1) z;1\right)
+2
G\left(1-z,1,-z^2+z+1;1\right)\\ +G\left(1-z,(1-z)
z,-z^2+z+1;1\right)+G\left(1-z,z^2-z+1,(z-1) z;1\right)
\\ -2
G\left(z,1,-z^2+z+1;1\right)-G\left(z,(1-z)
z,-z^2+z+1;1\right)-G\left(z,z^2-z+1,(z-1) z;1\right)
\\
+G((z-1) z;1) (2
G(1-z,1;1)+G(1-z,(1-z) z;1)-2 G(z,1;1)-G(z,(1-z) z;1))
\\
+2 G(1-z,0,(z-1) z;1)-2
G(z,0,(z-1) z;1)\Bigg]\, ,
\end{multline}
where $z$ is defined in Appendix \ref{notation} and we have automatically normalized all $G$ functions using the condition $G(a_1,...,a_n ; x) = G(a_1/x,...,a_n/x ; 1), ~ a_n \neq 0 $. Note, that the integrand depends on $t$ only through the combinations $y^2 = 1/t\bar{t}$, $\bar{t} =1-t$. Next, we can use the same procedure as above  and rewrite all MPLs in the canonical form with respect to the variable $z$. As almost all MPLs in the current expression depend only on variable $z$\footnote{The only exception is with MPLs $G\left(0;(z-1) z/y^2\right)$ and $G\left(0;-1/y^2\right)$. The latter are simple logarithms, which rewriting is easy.} the determination of integration constants for example with PSLQ is particularly simple. For example, we have
\begin{multline}
G\left(1-z,z^2-z+1;1\right)=-i \pi 
G\left(c_-;z\right)+G\left(0,c_-;z\right)-G\left(c_-,0;z\right)+G\left(c_-,1;z\right)
\\
-i \pi 
G\left(c_+;z\right)+G\left(0,c_+;z\right)-G\left(c_+,0;z\right)+G\left(c_+,1;z\right)
\\
+3 i \pi  G(0;z)+G(0,0;z)-3 G(0,1;z)-\frac{\pi ^2}{2}
\end{multline}
where $c_{\pm}= \frac{1}{2}(1 \pm \sqrt{3}i)$. Performing this rewriting and using notation from Appendix \ref{notation} we finally get
\begin{multline}
I_{1,1,1,1,1,1}^{v(0)}=\sum\limits_{i= \pm} \Big[ 6 J\left(M_2^1,\omega _0^z,\omega _{c_i}^z,\omega _0^z;s\right)+6 J\left(M_2^1,\omega _0^z,\omega _{c_i}^z,\omega
_1^z;s\right)-6 J\left(M_2^1,\omega _1^z,\omega _{c_i}^z,\omega _0^z;s\right)
\\
-6 J\left(M_2^1,\omega _1^z,\omega
_{c_i}^z,\omega _1^z;s\right)-10 J\left(M_2^1,\omega _0^z,\omega _0^z,\omega _{r_i}^z;s\right)-10 J\left(M_2^1,\omega
_0^z,\omega _1^z,\omega _{r_i}^z;s\right)
\\
+10 J\left(M_2^1,\omega _1^z,\omega _0^z,\omega _{r_i}^z;s\right)+10
J\left(M_2^1,\omega _1^z,\omega _1^z,\omega _{r_i}^z;s\right)\Big]+4 J\left(M_2^1,\omega _0^z,\omega _0^z,\omega _0^z;s\right)
\\
+4 J\left(M_2^1,\omega _0^z,\omega
_0^z,\omega _1^z;s\right)+4 J\left(M_2^1,\omega _0^z,\omega _1^z,\omega _0^z;s\right)+4 J\left(M_2^1,\omega
_0^z,\omega _1^z,\omega _1^z;s\right)
\\
-4 J\left(M_2^1,\omega _1^z,\omega _0^z,\omega _0^z;s\right)-4
J\left(M_2^1,\omega _1^z,\omega _0^z,\omega _1^z;s\right)-4 J\left(M_2^1,\omega _1^z,\omega _1^z,\omega
_0^z;s\right)
\\
-4 J\left(M_2^1,\omega _1^z,\omega _1^z,\omega _1^z;s\right)
-\frac{1}{3} \pi ^2 J\left(M_2^1,\omega _0^z;s\right)+\frac{1}{3} \pi ^2 J\left(M_2^1,\omega
_1^z;s\right)-12 \zeta (3) J\left(M_2^1;s\right)
\end{multline}
where $r_{\pm} = \frac{1}{2}(1\pm \sqrt{5})$. 
The obtained expression can be easily computed numerically. First, each $G$ function contained in each individual $J$ function can be easily evaluated with a prescribed precision using existing GiNaC library \cite{MPLs-numerics}. Then the final convergent integration over the variable $y$  can be taken numerically using standard methods. The result of this integration and its comparison with the sector decomposition method \cite{Binoth:2000ps,Binoth:2003ak,Binoth:2004jv,Heinrich:2008si,Bogner:2007cr,Bogner:2008ry,Kaneko:2009qx} can be found in Fig. \ref{figs:ep0M1}. 

\begin{figure}[t]
	\center{\includegraphics[width=0.7\textwidth]{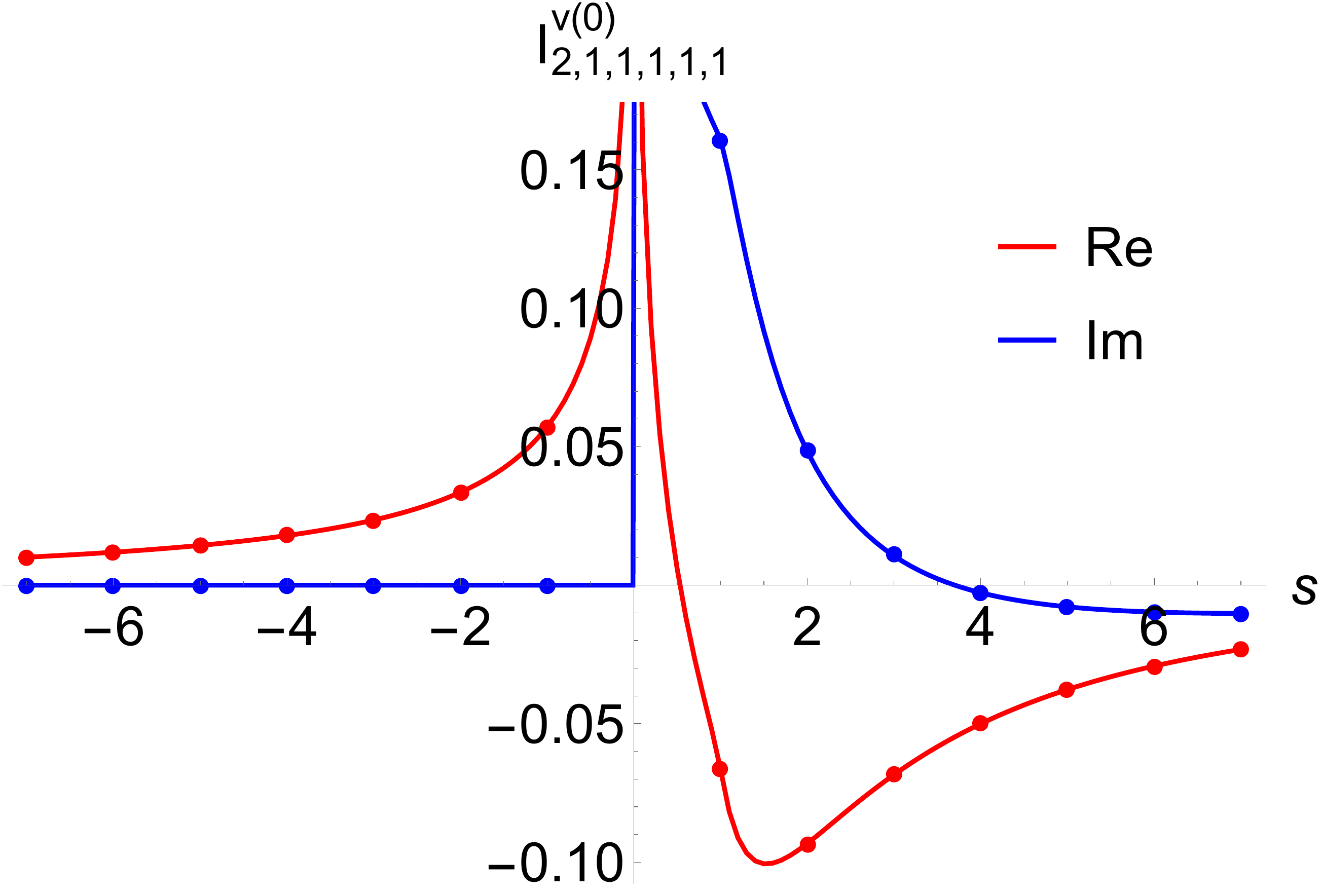}}
	\caption{Plot of the $\varepsilon^0$ correction to the $I^{v}_{2,1,1,1,1,1}$. The solid points represent values computed numerically with the FIESTA package \cite{Fiesta4}.}
	\label{ep0M2}
\end{figure}

The evaluation of $I^{v}_{2,1,1,1,1,1}$ master integral goes along the same lines and for leading order $\varepsilon$ correction we get
\begin{multline}
I_{2,1,1,1,1,1}^{v(0)}= \sum\limits_{i= \pm} \Big[-\frac{3}{2} J\left(M_2^2,\omega _{c_i}^z,\omega _0^z;s\right)-\frac{3}{2} J\left(M_2^2,\omega _{c_i}^z,\omega
_1^z;s\right)-3 J\left(M_3^3,\omega _0^z,\omega _{c_i}^z,\omega _0^z;s\right)
\\ 
-3 J\left(M_3^3,\omega _0^z,\omega
_{c_i}^z,\omega _1^z;s\right)
+3 J\left(M_3^3,\omega _1^z,\omega _{c_i}^z,\omega _0^z;s\right)+3 J\left(M_3^3,\omega
_1^z,\omega _{c_i}^z,\omega _1^z;s\right) 
 \\ 
+\frac{5}{2} J\left(M_2^2,\omega _0^z,\omega _{r_i}^z;s\right) +\frac{5}{2}
J\left(M_2^2,\omega _1^z,\omega _{r_i}^z;s\right)+5 J\left(M_3^3,\omega _0^z,\omega _0^z,\omega _{r_i}^z;s\right)
\\
 +5 J\left(M_3^3,\omega _0^z,\omega _1^z,\omega _{r_i}^z;s\right)-5 J\left(M_3^3,\omega _1^z,\omega _0^z,\omega
_{r_i}^z;s\right)-5 J\left(M_3^3,\omega _1^z,\omega _1^z,\omega _{r_i}^z;s\right)\Big]
\\
+2 J\left(M_3^3,\omega _1^z,\omega _0^z,\omega _1^z;s\right)
+2
J\left(M_3^3,\omega _1^z,\omega _1^z,\omega _0^z;s\right)+2 J\left(M_3^3,\omega _1^z,\omega _1^z,\omega
_1^z;s\right)
\\
-2 J\left(M_3^3,\omega _0^z,\omega _0^z,\omega _1^z;s\right)-2 J\left(M_3^3,\omega _0^z,\omega
_1^z,\omega _0^z;s\right)-2 J\left(M_3^3,\omega _0^z,\omega _1^z,\omega _1^z;s\right)
\\
-2 J\left(M_3^3,\omega _0^z,\omega _0^z,\omega
_0^z;s\right)+2 J\left(M_3^3,\omega
_1^z,\omega _0^z,\omega _0^z;s\right) -J\left(M_2^2,\omega _0^z,\omega _1^z;s\right)
\\
-J\left(M_2^2,\omega _1^z,\omega
_0^z;s\right)-
J\left(M_2^2,\omega _1^z,\omega _1^z;s\right)+
\frac{1}{12} \pi ^2 J\left(M_2^2;s\right)+\frac{1}{6}
\pi ^2 J\left(M_3^3,\omega _0^z;s\right)
\\
-\frac{1}{6} \pi ^2 J\left(M_3^3,\omega _1^z;s\right)-J\left(M_2^2,\omega
_0^z,\omega _0^z;s\right)+6 \zeta (3) J\left(M_3^3;s\right)
\end{multline}
The result of numerical evaluation and its comparison with the sector decomposition method can be found in Fig. \ref{ep0M2}. 

Using direct integration we can also find higher order $\varepsilon$ corrections for the first and second master integrals. For example, in the accompanying {\it Mathematica} notebook the reader can find expressions for $\varepsilon^1$ corrections.

\section{Analytical Frobenius solution}\label{Frobenius}

The integral solution of differential equations and direct integration of parametric representation considered above  are not the only methods one can use to find analytical solutions for Feynman integrals with elliptics. Instead of solving differential equations in terms of iterated integrals with algebraic kernels one can also try to find a power series solution. Moreover, it turns out that for non-planar vertex considered above this series solution will be exact in $\varepsilon$.    

\begin{figure}[t]
	\center{\includegraphics[width=0.89\textwidth]{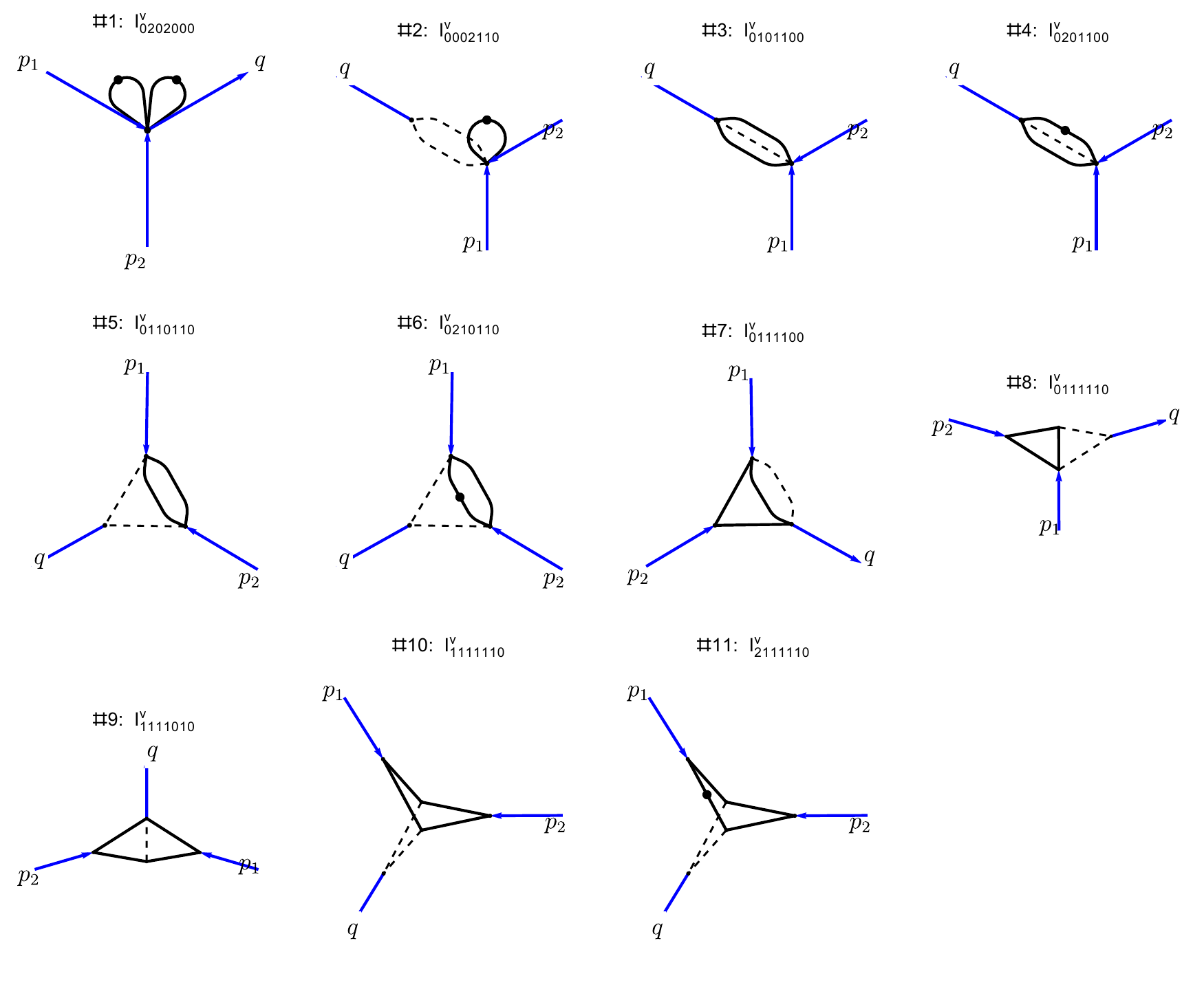}}
	\caption{IBP master integrals for integral family in Eq. \eqref{eqns:EV}. Solid lines stand for massive propagators and dashed for massless.}
	\label{figs:NonplanarVertexMasters}
\end{figure}

In this case we start with a set of 11 master integrals for integral family \eqref{eqns:EV}. The latter can be chosen as,  see Fig. \ref{figs:NonplanarVertexMasters} 
\begin{align}
{\bf I}_{\rm IBP} = \{&I^{v}_{0,2,0,2,0,0},& &I^{v}_{0,0,0,2,1,1},& &I^{v}_{0,1,0,1,1,0},& &I^{v}_{0,2,0,1,1,0},& &I^{v}_{0,1,1,0,1,1},& &I^{v}_{0,2,1,0,1,1}, \nonumber \\ &I^{v}_{0,1,1,1,1,0},& &I^{v}_{0,1,1,1,1,1},& &I^{v}_{1,1,1,1,0,1}, & &I^{v}_{1,1,1,1,1,1},&  &I^{v}_{2,1,1,1,1,1}\}^{\top}
\label{eqns:IBPmasters}
\end{align}

Here, we again use differentiation with respect to $s$ and IBP relations to write down a differential system they satisfy. In principle, we could try to find a series solution for this system directly. However, it is more convenient to first reduce this system to a simpler form, for example so called $A+\varepsilon B$ form. The matrix entries in this form are linear in regularization parameter $\varepsilon$, but the matrix itself is not necessary Fuchsian in the kinematic variable $s$. The basis integrals for this form, which we will also call canonical, where found to be given by
\begin{align}
J_1 & = \varepsilon^2 I^{v}_{0,2,0,2,0,0} & J_2  & = \varepsilon^2(-1+2\varepsilon) I^{v}_{0,0,0,2,1,1} \nonumber \\
J_3 & = \varepsilon^3(-2+3\varepsilon) I^{v}_{0,1,0,1,1,0} & J_4 & = \varepsilon^3 I^{v}_{0,2,0,1,1,0} \nonumber \\
J_5 & = \varepsilon^3(-1+2\varepsilon) I^{v}_{0,1,1,0,1,1} & J_6 & = \varepsilon^3 I^{v}_{0,2,1,0,1,1} \nonumber \\
J_7 & = \frac{\varepsilon^4}{4(1-3\varepsilon+2\varepsilon^2)} I^{v}_{0,2,0,2,0,0} + \varepsilon^4 s I^{v}_{0,1,1,1,1,0} & J_8 & = \varepsilon^4s I^{v}_{0,1,1,1,1,1} \\
J_9 & = \varepsilon^4 s I^{v}_{1,1,1,1,0,1} & J_{10} & = \varepsilon^3(1+2\varepsilon) s^2 I^{v}_{1,1,1,1,1,1} \nonumber \\
J_{11} & = \varepsilon^3 s^2(s+4) I^{v}_{2,1,1,1,1,1} \nonumber \label{eqns:canonicaltoIBPmasters}
\end{align}  
and the differential system in this basis takes the form
\begin{equation}
\label{eqns:diffeqn-s}
\frac{d {\bf J}}{ds} = {\bf M}\cdot {\bf J}\, ,
\end{equation}
where 
\begin{equation}
{\bf M} = \frac{\mathcal{M}_0}{s}+\frac{\mathcal{M}_1}{s-1}+\frac{\mathcal{M}_{-1}}{s+1}+\frac{\mathcal{M}_4}{s+4}+\mathcal{M}_0^1+\mathcal{M}_0^2s.
\end{equation}
and coefficient matrices $\mathcal{M}_i$ and $\mathcal{M}_i^j$ are given by
\begin{equation}
\mathcal{M}_0 = \left(
\begin{array}{ccccccccccc}
0 & 0 & 0 & 0 & 0 & 0 & 0 & 0 & 0 & 0 & 0 \\
0 & -\varepsilon  & 0 & 0 & 0 & 0 & 0 & 0 & 0 & 0 & 0 \\
0 & 0 & 1-2 \varepsilon  & 6 \varepsilon -4 & 0 & 0 & 0 & 0 & 0 & 0 & 0 \\
\frac{\varepsilon }{4} & 0 & \frac{1-2 \varepsilon}{4}  & \frac{4
	\varepsilon -3}{2} & 0 & 0 & 0 & 0 & 0 & 0 & 0 \\
0 & 0 & 0 & 0 & -2 \varepsilon  & 4 \varepsilon -2 & 0 & 0 & 0 & 0 & 0 \\
-\frac{\varepsilon }{8} & -\frac{\varepsilon }{4} & 0 & 0 & \frac{1-3
	\varepsilon}{4} & \frac{4 \varepsilon -3}{2} & 0 & 0 & 0 & 0 & 0 \\
0 & 0 & 0 & 0 & 0 & 0 & 0 & 0 & 0 & 0 & 0 \\
0 & 0 & -\frac{\varepsilon }{8} & \frac{\varepsilon }{4} & 0 & 0 & \frac{1-\varepsilon
}{2} & -\varepsilon  & 0 & 0 & 0 \\
0 & 0 & \frac{\varepsilon }{4} & -\frac{\varepsilon }{2} & 0 & 0 & \varepsilon -1 & 0 &
0 & 0 & 0 \\
0 & 0 & 0 & 0 & 0 & 0 & 0 & 0 & 0 & -2 \varepsilon  & 2 \varepsilon +1 \\
0 & 0 & 0 & 0 & 0 & 0 & 0 & 0 & 0 & -\frac{4 \varepsilon + 1}{4}  & \varepsilon +1 \\
\end{array}
\right),
\end{equation}
\begin{equation}
\mathcal{M}_1 = \left(
\begin{array}{ccccccccccc}
0 & 0 & 0 & 0 & 0 & 0 & 0 & 0 & 0 & 0 & 0 \\
0 & 0 & 0 & 0 & 0 & 0 & 0 & 0 & 0 & 0 & 0 \\
0 & 0 & 0 & 0 & 0 & 0 & 0 & 0 & 0 & 0 & 0 \\
-\frac{\varepsilon }{4} & 0 & \frac{2 \varepsilon -1}{4} & -\frac{6
	\varepsilon -3}{2}  & 0 & 0 & 0 & 0 & 0 & 0 & 0 \\
0 & 0 & 0 & 0 & 0 & 0 & 0 & 0 & 0 & 0 & 0 \\
0 & 0 & 0 & 0 & 0 & 0 & 0 & 0 & 0 & 0 & 0 \\
0 & 0 & 0 & 0 & 0 & 0 & 0 & 0 & 0 & 0 & 0 \\
0 & 0 & 0 & 0 & 0 & 0 & 0 & 0 & 0 & 0 & 0 \\
0 & 0 & 0 & 0 & 0 & 0 & 0 & 0 & 0 & 0 & 0 \\
0 & 0 & 0 & 0 & 0 & 0 & 0 & 0 & 0 & 0 & 0 \\
\frac{5 \varepsilon }{64} & 0 & -\frac{10 \varepsilon -5}{64}  & \frac{30
	\varepsilon -15}{32}  & 0 & 0 & 0 & 0 & 0 & 0 & 0 \\
\end{array}
\right),
\end{equation}

\begin{equation}
\mathcal{M}_{-1} = \left(
\begin{array}{ccccccccccc}
0 & 0 & 0 & 0 & 0 & 0 & 0 & 0 & 0 & 0 & 0 \\
0 & 0 & 0 & 0 & 0 & 0 & 0 & 0 & 0 & 0 & 0 \\
0 & 0 & 0 & 0 & 0 & 0 & 0 & 0 & 0 & 0 & 0 \\
0 & 0 & 0 & 0 & 0 & 0 & 0 & 0 & 0 & 0 & 0 \\
0 & 0 & 0 & 0 & 0 & 0 & 0 & 0 & 0 & 0 & 0 \\
\frac{\varepsilon }{8} & \frac{\varepsilon }{4} & 0 & 0 & \frac{3 \varepsilon
	-1}{4} & \frac{1-4 \varepsilon }{2}  & 0 & 0 & 0 & 0 & 0 \\
0 & 0 & 0 & 0 & 0 & 0 & 0 & 0 & 0 & 0 & 0 \\
0 & 0 & 0 & 0 & 0 & 0 & 0 & 0 & 0 & 0 & 0 \\
0 & 0 & 0 & 0 & 0 & 0 & 0 & 0 & 0 & 0 & 0 \\
0 & 0 & 0 & 0 & 0 & 0 & 0 & 0 & 0 & 0 & 0 \\
\frac{3 \varepsilon }{32} & \frac{3 \varepsilon }{16} & 0 & 0 & \frac{9
	\varepsilon -3}{16} & -\frac{12 \varepsilon -3}{8}  & 0 & 0 & 0 & 0 & 0 \\
\end{array}
\right),
\end{equation}

\begin{equation}
\mathcal{M}_{4} = \left(
\begin{array}{ccccccccccc}
0 & 0 & 0 & 0 & 0 & 0 & 0 & 0 & 0 & 0 & 0 \\
0 & 0 & 0 & 0 & 0 & 0 & 0 & 0 & 0 & 0 & 0 \\
0 & 0 & 0 & 0 & 0 & 0 & 0 & 0 & 0 & 0 & 0 \\
0 & 0 & 0 & 0 & 0 & 0 & 0 & 0 & 0 & 0 & 0 \\
0 & 0 & 0 & 0 & 0 & 0 & 0 & 0 & 0 & 0 & 0 \\
0 & 0 & 0 & 0 & 0 & 0 & 0 & 0 & 0 & 0 & 0 \\
0 & 0 & 0 & 0 & 0 & 0 & 0 & 0 & 0 & 0 & 0 \\
0 & 0 & 0 & 0 & 0 & 0 & 0 & 0 & 0 & 0 & 0 \\
0 & 0 & 0 & 0 & 0 & 0 & 0 & 0 & 0 & 0 & 0 \\
0 & 0 & 0 & 0 & 0 & 0 & 0 & 0 & 0 & 0 & -2 \varepsilon -1 \\
0 & 0 & 0 & 0 & 0 & 0 & 0 & 0 & 0 & 0 & -\varepsilon  \\
\end{array}
\right),
\end{equation}

\begin{equation}
\mathcal{M}_{0}^1 =\left(
\begin{array}{ccccccccccc}
0 & 0 & 0 & 0 & 0 & 0 & 0 & 0 & 0 & 0 & 0 \\
0 & 0 & 0 & 0 & 0 & 0 & 0 & 0 & 0 & 0 & 0 \\
0 & 0 & 0 & 0 & 0 & 0 & 0 & 0 & 0 & 0 & 0 \\
0 & 0 & 0 & 0 & 0 & 0 & 0 & 0 & 0 & 0 & 0 \\
0 & 0 & 0 & 0 & 0 & 0 & 0 & 0 & 0 & 0 & 0 \\
0 & 0 & 0 & 0 & 0 & 0 & 0 & 0 & 0 & 0 & 0 \\
0 & 0 & 0 & \varepsilon  & 0 & 0 & 0 & 0 & 0 & 0 & 0 \\
0 & 0 & 0 & -\frac{\varepsilon }{2} & 0 & \varepsilon  & 0 & 0 & 0 & 0 & 0 \\
0 & 0 & 0 & \varepsilon  & 0 & 0 & 0 & 0 & 0 & 0 & 0 \\
0 & 0 & 0 & 0 & 0 & 0 & 0 & 0 & 0 & 0 & 0 \\
\frac{\varepsilon }{16} & -\frac{3 \varepsilon }{16} & -\frac{\varepsilon }{4} &
\frac{18 \varepsilon -5}{16} & -\frac{9 \varepsilon -3}{16}  & \frac{12
	\varepsilon -3}{8} & 1-\varepsilon  & -\frac{\varepsilon }{2} & 0 & 0 & 0 \\
\end{array}
\right),
\end{equation}

\begin{equation}
\mathcal{M}_{0}^2 =\left(
\begin{array}{ccccccccccc}
0 & 0 & 0 & 0 & 0 & 0 & 0 & 0 & 0 & 0 & 0 \\
0 & 0 & 0 & 0 & 0 & 0 & 0 & 0 & 0 & 0 & 0 \\
0 & 0 & 0 & 0 & 0 & 0 & 0 & 0 & 0 & 0 & 0 \\
0 & 0 & 0 & 0 & 0 & 0 & 0 & 0 & 0 & 0 & 0 \\
0 & 0 & 0 & 0 & 0 & 0 & 0 & 0 & 0 & 0 & 0 \\
0 & 0 & 0 & 0 & 0 & 0 & 0 & 0 & 0 & 0 & 0 \\
0 & 0 & 0 & 0 & 0 & 0 & 0 & 0 & 0 & 0 & 0 \\
0 & 0 & 0 & 0 & 0 & 0 & 0 & 0 & 0 & 0 & 0 \\
0 & 0 & 0 & 0 & 0 & 0 & 0 & 0 & 0 & 0 & 0 \\
0 & 0 & 0 & 0 & 0 & 0 & 0 & 0 & 0 & 0 & 0 \\
0 & 0 & 0 & -\varepsilon  & 0 & \varepsilon  & 0 & 0 & 0 & 0 & 0 \\
\end{array}
\right).
\end{equation}
Now note, that master integrals $J_1$ and $J_2$ are easy and direct calculation of corresponding Feynman diagrams gives 
\begin{align}
J_1 &= \varepsilon ^2 \Gamma (\varepsilon )^2, \nonumber \\
J_2 &= \frac{4^{-\varepsilon } \varepsilon ^2 (2 \varepsilon -1) \Gamma (1-\varepsilon )^2 \Gamma (\varepsilon )^2 (-s)^{-\varepsilon }}{\Gamma (2-2 \varepsilon )}. \label{eqns:J1J2}
\end{align}
To find a Frobenius series solution for the rest of canonical master integrals we will use an ansatz
\begin{equation}
{\bf J} = \sum_{\lambda}\sum_{n=0}^{\infty} {\bf C}_{\lambda}(n) s^{n+\lambda}\, ,
\end{equation} 
Substituting this ansatz into differential system \eqref{eqns:diffeqn-s} we obtain a recurrence relations on the coefficient matrices ${\bf C}_{\lambda}(n)$ together with constraints on the set of $\lambda$'s. These matrix difference equations are then reduced to ordinary first-order inhomogeneous linear difference equations of the form:
\begin{equation}
\label{eqns:GenReqEq}
f(n+1)=H(n)f(n)+Q(n)
\end{equation}
This equation can be easily solved and the general solution has the form 
\begin{equation}
\label{eqns:GenReqSol}
f(n) = f_{\rm hom}(n)\left(\sum\limits_{k=l}^{n-1}\frac{Q(k)}{ f_{\rm hom}(k+1)} +C\right)\, ,\quad f_{\rm hom}(n) = \prod\limits_{m=l}^{n-1}H(m)\, ,
\end{equation}
where  $f_{\rm hom}(n)$ is the solution of corresponding homogeneous equation and $C$ is some constant to be determined from the boundary conditions\footnote{In general, $C$ is a periodic function of $n$ with period 1, but in our case, we can obviously ignore this fact and consider $C$ to be just a constant.}. The lower limit $l$ is chosen so that the sum in Eq. \eqref{eqns:GenReqSol} is finite. Now, let us consider Frobenius solution for canonical master integrals in detail.

\subsection{Non-elliptic canonical master integrals}

The non-elliptic canonical master integrals are given by integrals $J_i$ $(i=1,\ldots 9)$. Consider as an example calculation of canonical master integrals $J_5$ and $J_6$. The differential system for integrals $J_5$ and $J_6$ has the form
\begin{align}
\label{eqns:J5eqPrel}
\frac{d J_5}{ds} & =  - \frac{2\varepsilon J_5 }{s}  -  \frac{2(1-2\varepsilon)J_6}{s},  \\
\frac{d J_6}{ds} & =  -  \frac{\varepsilon J_1}{8s(s+1)}  -  \frac{\varepsilon J_2}{4s(s+1)}  + \frac{(1-3\varepsilon)J_5}{4s(s+1)}  -  \frac{(3+2s-4\varepsilon)J_6}{2s(s+1)}. 
\end{align}
Differentiating the first of these equations and substituting into it the expression for $J_6$ and $d J_6/ds$ obtained from the original system we get the following second-order differential equation for $J_5$:
\begin{equation}
\label{eqns:J5DEExact}
s^2(1+s)\frac{d^2 J_5}{ds^2} + \frac{s(5+4s(1+\varepsilon))}{2}\frac{d J_5}{ds} + \frac{1+(1+4s)\varepsilon-2\varepsilon^2}{2}J_5 - \frac{\varepsilon(1-2\varepsilon)}{4}\left(J_1 + 2 J_2\right) = 0. 
\end{equation}
To find a homogeneous solution of Eq. \eqref{eqns:J5DEExact} we substitute into its homogeneous part  an ansatz
\begin{equation}
J_5^{\rm hom}(s) = \sum_{n=0}^{\infty}a_n^{(\lambda )} s^{\lambda + n}
\end{equation}
and equate coefficients in front of $s^{\lambda + n}$. This way we obtain recurrence relation for $a_n^{(\lambda )}$ coefficients:
\begin{equation}
2 (\lambda +n) (\lambda +n+2 \varepsilon -1)  a_{n-1}^{(\lambda )} +  (\lambda +n-\varepsilon +1) (2 \lambda +2 n+2
\varepsilon +1) a_{n}^{(\lambda )} = 0\, .
\end{equation}
Its solution is simple and we get (see Eq. \eqref{eqns:GenReqSol} )
\begin{equation}
a_n^{(\lambda )} = \frac{(-1)^{n+1} \Gamma (n+\lambda +1) \Gamma (-\varepsilon +\lambda +3) \Gamma \left(\varepsilon +\lambda
	+\frac{5}{2}\right) \Gamma (n+2 \varepsilon +\lambda )}{\Gamma (\lambda +2) \Gamma (2 \varepsilon +\lambda +1)
	\Gamma (n-\varepsilon +\lambda +2) \Gamma \left(n+\varepsilon +\lambda +\frac{3}{2}\right)}\, .
\end{equation}
Additionally, to find exponents $\lambda$ for two homogeneous solutions we consider the constraint at $s\to 0$:
\begin{equation}
a_0^{(\lambda )} (\lambda -\varepsilon +1) (2 \lambda +2 \varepsilon +1) = 0\, ,
\end{equation}
so that general homogeneous solution is given by
\begin{equation}
J_5^{\rm hom}(s) = C_1 \sum_{n=0}^{\infty}a_n^{(-1/2-\varepsilon)} s^{-1/2-\varepsilon + n} + 
 C_2 \sum_{n=0}^{\infty}a_n^{(-1+\varepsilon)} s^{-1+\varepsilon + n}
\end{equation}
The particular solution for Eq. \eqref{eqns:J5DEExact} is found similarly with the ansatz 
\begin{equation}
J_5^{\rm nonhom}(s) = \sum_{n=0}^{\infty}(b_n s^{n} + c_n s^{n-\varepsilon})\, ,
\end{equation}
where the shifts of $s$-exponents from $n$ follow $s$-exponents of master integrals $J_1$ and $J_2$ \eqref{eqns:J1J2}. The recurrence relations for coefficients $b_n$ and $c_n$ are found to be 
\begin{align}
b_{n+1} & = \frac{2(n+1)(n+2\varepsilon)}{(n+2-\varepsilon)(2n+3+2\varepsilon)}b_n, \\ 
c_{n+1} & = \frac{2(n-\varepsilon+1)(n+\varepsilon)}{(3+2n)(n-2\varepsilon+2)} c_n .
\end{align} 
with solutions
\begin{equation}
\label{eqns:bsolGenJ5}
b_n =
\frac{C_3 (-1)^{n+1} \Gamma (3-\varepsilon ) \Gamma \left(\varepsilon +\frac{5}{2}\right) \Gamma
	(n+1) \Gamma (n+2 \varepsilon )}{\Gamma (2 \varepsilon +1) \Gamma (n-\varepsilon +2) \Gamma
	\left(n+\varepsilon +\frac{3}{2}\right)},
\end{equation}
and
\begin{equation}
\label{eqns:csolGenJ5}
c_n = -\frac{3 C_4 (-1)^n \sin (\pi  \varepsilon ) \Gamma (2-2 \varepsilon ) \Gamma (n-\varepsilon +1)
	\Gamma (n+\varepsilon )}{2 \sqrt{\pi } \varepsilon  \Gamma \left(n+\frac{3}{2}\right) \Gamma (n-2
	\varepsilon +2)}
\end{equation}
Finally, the constants $C_i$ $(i=1,\ldots 4)$ are found from boundary conditions at $s\to 0$ and we get  
\begin{multline}
\label{eqns:J5exactSol}
\frac{2^{2 \varepsilon +1} \sin (\pi  \varepsilon )}{\pi ^{3/2} \varepsilon ^3 (2 \varepsilon -1)} J_5 = 
\sum_n\frac{(-1)^{n-\varepsilon} s^{n-\varepsilon } \Gamma (n-\varepsilon +1) \Gamma (n+\varepsilon
	)}{\Gamma \left(n+\frac{3}{2}\right) \Gamma (n-2 \varepsilon +2)}\\+\sum_n\frac{(-1)^{n+1} s^n
	\Gamma (n+1) \Gamma (n+2 \varepsilon )}{\Gamma (n-\varepsilon +2) \Gamma \left(n+\varepsilon
	+\frac{3}{2}\right)}\, .
\end{multline}
where $\sum\limits_n = \sum_{n=0}^\infty$ and both sums are convergent at $|s| \leq 1$.
Next, the solution for $J_6$ integral is obtained simply by substituting solution \eqref{eqns:J5exactSol} into Eq. \eqref{eqns:J5eqPrel}
\begin{multline}
\label{eqns:J6exactSol}
\frac{2^{2 (\varepsilon +1)} \sin (\pi  \varepsilon )}{\pi ^{3/2} \varepsilon ^3} J_6 = 
\sum_n\frac{(-1)^{n-\varepsilon} s^{n-\varepsilon } \Gamma (n-\varepsilon +1) \Gamma (n+\varepsilon
	+1)}{\Gamma \left(n+\frac{3}{2}\right) \Gamma (n-2 \varepsilon +2)}\\-\sum_n\frac{(-1)^{n} s^n
	\Gamma (n+1) \Gamma (n+2 \varepsilon +1)}{\Gamma (n-\varepsilon +2) \Gamma \left(n+\varepsilon
	+\frac{3}{2}\right)}\, .
\end{multline}
In the same way we can obtain expressions for all other non-elliptic master integrals:
\begin{align}
\frac{2^{2 \varepsilon -1} \sin (\pi  \varepsilon ) \Gamma (\varepsilon )}{\pi ^{3/2} \varepsilon ^3 (3 \varepsilon -2)} J_3 =& \sum_n\frac{s^n \Gamma (n+\varepsilon ) \Gamma (n+2 \varepsilon -1)}{\Gamma (n-\varepsilon +2) \Gamma
	\left(n+\varepsilon +\frac{1}{2}\right)}\, , \\
\frac{4^{\varepsilon } \sin (\pi  \varepsilon ) \Gamma (\varepsilon )}{\pi ^{3/2} \varepsilon ^3} J_4 =& 
\sum_n\frac{s^n \Gamma (n+\varepsilon ) \Gamma (n+2 \varepsilon )}{\Gamma (n-\varepsilon +2) \Gamma
	\left(n+\varepsilon +\frac{1}{2}\right)}\, , \\
\frac{4^{\varepsilon } \sin (\pi  \varepsilon ) \Gamma (\varepsilon )}{\pi ^{3/2} \varepsilon ^4} J_7 =&
\sum_n\frac{s^{n+1} \Gamma (n+\varepsilon ) \Gamma (n+2 \varepsilon )}{(n+1) \Gamma (n-\varepsilon +2) \Gamma
	\left(n+\varepsilon +\frac{1}{2}\right)} +\frac{4^{\varepsilon -1} \sin (\pi  \varepsilon ) \Gamma (\varepsilon
	)^3}{\pi ^{3/2} (\varepsilon -1) (2 \varepsilon -1)}\, , \\
\frac{2^{2 \varepsilon +2} \sin (\pi  \varepsilon ) \Gamma (\varepsilon )}{\pi ^{3/2} \varepsilon ^4} J_8 =&  \Gamma (\varepsilon ) \sum_n\frac{(-1)^{n-\varepsilon} s^{n-\varepsilon +1} \Gamma (n-\varepsilon +1)
	\Gamma (n+\varepsilon +1)}{(n+1) \Gamma \left(n+\frac{3}{2}\right) \Gamma (n-2 \varepsilon
	+2)}\nonumber \\ & -\sum_n\frac{s^{n+1} \left((-1)^n \Gamma (n+2) \Gamma (\varepsilon )+\Gamma (n+\varepsilon
	+1)\right) \Gamma (n+2 \varepsilon +1)}{(n+1) (n+\varepsilon +1) \Gamma (n-\varepsilon +2) \Gamma \left(n+\varepsilon
	+\frac{3}{2}\right)}\, , \\
\frac{2^{2 \varepsilon +1} \sin (\pi  \varepsilon ) \Gamma (\varepsilon )}{\pi ^{3/2} \varepsilon ^4} J_9 =& \sum_n\frac{s^{n+1} \Gamma (n+\varepsilon +1) \Gamma (n+2 \varepsilon +1)}{(n+1)^2 \Gamma (n-\varepsilon +2)
	\Gamma \left(n+\varepsilon +\frac{3}{2}\right)}\, \, 
\end{align}
where again all the present sums are convergent in the interval $|s| \leq 1$. These 1-dimensional sums can be further rewritten in terms of generalized hypergeometric functions as
\begin{align}
 \frac{2^{2 \varepsilon -1} \sin (\pi  \varepsilon ) \Gamma (\varepsilon )}{\pi ^{3/2} \varepsilon ^3 (3 \varepsilon -2)} J_3 =& \frac{\Gamma(\ep)\Gamma(2\ep-1)}{\Gamma(2-\ep)\Gamma(\ep+1/2)}\pFq{3}{2}{2\ep-1,\ep,1}{2-\ep,1/2+\ep}{s} , \\
\frac{4^{\varepsilon } \sin (\pi  \varepsilon ) \Gamma (\varepsilon )}{\pi ^{3/2} \varepsilon ^3} J_4 =& 
\frac{\Gamma(\ep)\Gamma(2\ep)}{\Gamma(2-\ep)\Gamma(\ep+1/2)}\pFq{3}{2}{2\ep,\ep,1}{2-\ep,1/2+\ep}{s},
\\
\frac{2^{2 \varepsilon +1} \sin (\pi  \varepsilon )}{\pi ^{3/2} \varepsilon ^3 (2 \varepsilon -1)} J_5 = &
\frac{2\sqrt{\pi}(-s)^{-\ep}\csc(\pi \ep)}{\Gamma(2-2\ep)}\pFq{3}{2}{1-\ep,\ep,1}{2-2\ep,3/2}{-s} \nonumber\\
&-\frac{\Gamma(2\ep)}{\Gamma(2-\ep)\Gamma(3/2+\ep)}\pFq{3}{2}{1,2\ep,1}{2-\ep,3/2+\ep}{-s},\\
\frac{2^{2 (\varepsilon +1)} \sin (\pi  \varepsilon )}{\pi ^{3/2} \varepsilon ^3} J_6 = &
\frac{2\sqrt{\pi}(-s)^{-\ep}\csc(\pi \ep)}{\Gamma(2-2\ep)}\pFq{3}{2}{1+\ep,1-\ep,1}{2-2\ep,3/2}{-s}\nonumber\\
&-\frac{\Gamma(1+2\ep)}{\Gamma(2-\ep)\Gamma(3/2+\ep)}\pFq{3}{2}{1,1+2\ep,1}{2-\ep,3/2+\ep}{-s},\\
\frac{4^{\varepsilon } \sin (\pi  \varepsilon ) \Gamma (\varepsilon )}{\pi ^{3/2} \varepsilon ^4} J_7 =&
\frac{\Gamma(\ep)\Gamma(2\ep)}{\Gamma(2-\ep)\Gamma(\ep+1/2)}\pFq{4}{3}{2\ep,\ep,1,1}{2-\ep,1/2+\ep,2}{s}+\frac{4^{\varepsilon -1} \sin (\pi  \varepsilon ) \Gamma (\varepsilon
)^3}{\pi ^{3/2} (\varepsilon -1) (2 \varepsilon -1)},\\
\frac{2^{2 \varepsilon +2} \sin (\pi  \varepsilon ) \Gamma (\varepsilon )}{\pi ^{3/2} \varepsilon ^4} J_8 =& 
\frac{2\sqrt{\pi}(-1)^{-\ep}s^{1-\ep}\Gamma(\ep+1)}{\Gamma(2-2\ep)}\pFq{4}{3}{1+\ep,1-\ep,1,1}{2-2\ep,3/2,2}{-s}
\nonumber\\
&-\frac{s\Gamma(\ep+1)\Gamma(1+2\ep)}{(1+\ep)\Gamma(2-\ep)\Gamma(3/2+\ep)}\pFq{5}{4}{1+\ep,1+\ep,1+2\ep,1,1}{2,3/2+\ep,2+\ep,2-\ep}{s}
\nonumber\\
&-\frac{s\Gamma(\ep)\Gamma(1+2\ep)}{(1+\ep)\Gamma(2-\ep)\Gamma(3/2+\ep)}\pFq{4}{3}{1+\ep,1+2\ep,1,1}{2-\ep,3/2+\ep,2+\ep}{-s},\\
\frac{2^{2 \varepsilon +1} \sin (\pi  \varepsilon ) \Gamma (\varepsilon )}{\pi ^{3/2} \varepsilon ^4} J_9  = &
\frac{s\Gamma(1+\ep)\Gamma(1+2\ep)}{\Gamma(2-\ep)\Gamma(3/2+\ep)}\pFq{5}{4}{1+\ep,1+2\ep,1,1,1}{2,2,3/2+\ep,2-\ep}{s}.
\end{align}

\subsection{Elliptic canonical master integrals}

The evaluation of series solutions for elliptic canonical master integrals $J_{10}$ and $J_{11}$ goes along the same lines as that for non-elliptic integrals. The only difference is extra preliminary manipulation with original differential equation to get results as compact as possible. The original second order differential equation for the integral $J_{10}$ has the form
\begin{equation}
\label{eqns:J10ExactEquation}
 s^2(4+s)\frac{d^2 J_{10}(s)}{ds^2} + s(s+2(2+s)\varepsilon)\frac{dJ_{10}(s)}{ds} + (1-2\varepsilon)J_{10}(s) + J^{\rm inhom}(s) = 0
\end{equation}
where the inhomogeneous part $J^{\rm inhom}(s)$ is given by
\begin{multline}
J^{\rm inhom}(s) = \frac{\varepsilon  (2 \varepsilon +1) (s (4 s+11)-5) s J_1}{16
	\left(s^2-1\right)}-\frac{3 \varepsilon  (2
	\varepsilon +1) s^2 J_2}{4 (s+1)}-\frac{(2 \varepsilon
	+1) s (2 \varepsilon  (8 s-3)-5) J_3}{16 (s-1)}
\\
-\frac{(2 \varepsilon +1) s (6 \varepsilon +2 s
	(2 \varepsilon  (8 s-17)+5)+5) J_4}{8 (s-1)}-\frac{3 \left(6 \varepsilon ^2+\varepsilon -1\right) s^2 J_5}{4 (s+1)}\\ +\frac{(2 \varepsilon +1) s^2 (4 \varepsilon  (2 s+5)-3)
	J_6}{2 (s+1)}
+4 \left(-2 \varepsilon ^2+\varepsilon +1\right) s J_7-2 \varepsilon  (2 \varepsilon +1) s J_8\, .
\label{eqns:J10EqInhomPart}
\end{multline}
The homogeneous part of this equation has the same structure as in non-elliptic case, that is the corresponding difference equation for power series coefficients are of  first order\footnote{Note that this property holds only for the integral $J_{10}$. If we write an equation for $J_{11}$ then this property will be lost.}. The only difference is in  inhomogeneous part $J^{\rm inhom}(s)$. This time the inhomogeneous part contains series in the variable $s$ and as follows from Eq. \eqref{eqns:GenReqSol} the solution will contain at least double sums. Another problem is the presence of rational factors such as $1/(s \pm 1)$ in the inhomogeneous part. There are several ways to treat this problem. The most obvious one is to use "brute force" and expand products of rational factors with lower master integrals in double series of the form   
\begin{equation}
\label{eqns:brute-force}
\frac{\sum\limits_{n=0}^{\infty}c_n s^n}{s-1} = -\sum\limits_{n=0}^{\infty}\sum\limits_{k=0}^{n}c_k s^n, \qquad \frac{\sum\limits_{n=0}^{\infty}c_n s^n}{s+1} = \sum\limits_{n=0}^{\infty}\sum\limits_{k=0}^{n}(-1)^{k+n}c_k s^n.
\end{equation}
This will lead however to the presence of 1-sums in the inhomogeneous part of difference equation for power coefficients. Hence, taking into account Eq. \eqref{eqns:GenReqSol}, we can immediately conclude that the solution for $J_{10}$ will also contain triple sums. This is extremely undesirable since then the answer will be unnecessarily cumbersome. Fortunately, there is a very elegant way out of this situation. We can use differential system \eqref{eqns:diffeqn-s} to eliminate master integrals $J_3$ and $J_5$ from the expression for inhomogeneous part \eqref{eqns:J10EqInhomPart}. This way for the latter we get   
\begin{multline}
J^{\rm inhom}(s) = -\frac{s\varepsilon(2\varepsilon-5)(1+2\varepsilon)}{8(2\varepsilon-1)}J_1 + \frac{s(1+2\varepsilon)(-5+\varepsilon(-1+6\varepsilon+16s(3\varepsilon-1)))}{4(2\varepsilon-1)}J_4 \\-s^2(1+2\varepsilon)(3+4\varepsilon)J_6 
-4s(1+\varepsilon-2\varepsilon^2)J_7 - 2s\varepsilon(1+2\varepsilon)J_8 \\ - \frac{s^2(1+2\varepsilon)(-5+2(8s-3)\varepsilon)}{4(1-2\varepsilon)}\frac{dJ_4}{ds} -3s^3(1+2\varepsilon)\frac{dJ_6}{ds}\, ,
\end{multline}
where instead of $J_3$ and $J_5$ we have $\frac{dJ_4}{ds}$ and $\frac{dJ_6}{ds}$. Surprisingly, such simple change leads to the disappearance of all bad rational factors and
makes the equation for $J_{10}$ much more simple. The power series solution for the latter will contain now at most double sums.

After this preparation step, the solution of Eq. \eqref{eqns:J10ExactEquation} goes along the same lines as in non-elliptic case. First, we make an ansatz\footnote{The contribution of homogeneous solution is zero in this case.} 
\begin{equation}
\label{anzatzJ10FinalGen}
J_{10} = \sum\limits_{n=2}^{\infty} \left(c_n s^n + d_n s^{n-\varepsilon}\right) 
\end{equation}
and substitute it into equation for $J_{10}$. Then, equating coefficients in front of $s^n$ and $s^{n-\varepsilon}$, we obtain two inhomogeneous first order difference equations for the coefficients $c_n$ and $d_n$:
\begin{multline}
\frac{n c(n) (2 \varepsilon +n)}{(2 n+1) (2 \varepsilon +2
	n+1)}+c(n+1)=\\=\frac{\pi ^{3/2} 2^{-2 \varepsilon -3} (2 \varepsilon +1)
	\varepsilon ^3 \csc (\pi  \varepsilon )(-1)^{n} \Gamma (\varepsilon ) (\varepsilon
	+3 n) \Gamma (n+1) \Gamma (n+2 \varepsilon +1)}{n (2 n+1) \Gamma (\varepsilon )
	(\varepsilon +n) \Gamma (n-\varepsilon +1) \Gamma \left(n+\varepsilon
	+\frac{3}{2}\right)}\\-\frac{\pi ^{3/2} 2^{-2 \varepsilon -3} (2 \varepsilon +1)
	\varepsilon ^3 \csc (\pi  \varepsilon )(3 \varepsilon +5 n) \Gamma (n+\varepsilon
	) \Gamma (n+2 \varepsilon +1)}{n (2 n+1) \Gamma (\varepsilon )
	(\varepsilon +n) \Gamma (n-\varepsilon +1) \Gamma \left(n+\varepsilon
	+\frac{3}{2}\right)},
\end{multline}
\begin{multline}
\frac{d(n) (n-\varepsilon ) (\varepsilon +n)}{(2 n+1) (-2
	\varepsilon +2 n+1)}+d(n+1)=\\=-\frac{\pi ^{3/2} 2^{-2 \varepsilon -3} (2 \varepsilon
	+1) \varepsilon ^3 \csc (\pi  \varepsilon ) (-1)^{n-\varepsilon } (3 n-2
	\varepsilon ) \Gamma (n-\varepsilon ) \Gamma (n+\varepsilon +1)}{n (-2
	\varepsilon +2 n+1) \Gamma \left(n+\frac{3}{2}\right) \Gamma (n-2
	\varepsilon +1)}=0.
\end{multline}
The latter are then easily solved (see Eq. \eqref{eqns:GenReqSol}) with arbitrary coefficients fixed from boundary conditions at $s\to 0$. Putting all together, for $J_{10}$ master integral we get a relatively simple expression\footnote{Note, that using "brute force" \eqref{eqns:brute-force} leads to expressions two orders of magnitude bigger in size.}: 
\begin{multline}
\frac{J_{10} \sin (\pi  \varepsilon ) }{s^2\pi^{3/2} \varepsilon^3(1+2\varepsilon)} = \\ \sum\limits_{j,n}\frac{ (-2 \varepsilon +3 j+3) (-1)^{n-\varepsilon } 4^{-\varepsilon +j-n-2} \Gamma \left(j-\varepsilon
		+\frac{3}{2}\right) \Gamma (n-\varepsilon +2) \Gamma (n+\varepsilon +2) s^{n-\varepsilon }}{(j+1) (-\varepsilon
		+j+1) \Gamma \left(n+\frac{5}{2}\right) \Gamma (j-2 \varepsilon +2) \Gamma \left(n-\varepsilon
		+\frac{5}{2}\right)}
\\
+\sum\limits_{j,n}\frac{ (\varepsilon +3 j+6)(-1)^{n+1}  4^{-\varepsilon +j-n-1} \Gamma
		\left(j+\frac{5}{2}\right) \Gamma (n+2) \Gamma (n+2 \varepsilon +2)s^n}{(j+2) (\varepsilon +j+2) \Gamma
		\left(n+\frac{5}{2}\right) \Gamma (j-\varepsilon +3) \Gamma \left(n+\varepsilon
		+\frac{5}{2}\right)}
\\
+\sum\limits_{j,n}\frac{ (3 \varepsilon +5 j+10) (-1)^{-j+n} 4^{-\varepsilon +j-n-1}
		\Gamma \left(j+\frac{5}{2}\right) \Gamma (n+2) \Gamma (j+\varepsilon +2)  \Gamma (n+2 \varepsilon +2) s^n}{(j+2)
		\Gamma (\varepsilon ) (\varepsilon +j+2) \Gamma (j+3) \Gamma \left(n+\frac{5}{2}\right) \Gamma (j-\varepsilon +3) \Gamma
		\left(n+\varepsilon +\frac{5}{2}\right)}
\\
-\sum\limits_{n}\frac{3 \pi ^{1/2} (\varepsilon +1) (-1)^n 
		2^{-2 \varepsilon -2 n-5}\Gamma (n+2) \Gamma (n+2 \varepsilon +2) s^n }{\Gamma (2-\varepsilon ) \Gamma \left(n+\frac{5}{2}\right) \Gamma
		\left(n+\varepsilon +\frac{5}{2}\right)}
\\
+\sum\limits_{n}\frac{ 2^{-2 \varepsilon -2} (-1)^n  (\varepsilon +3 n+6)  \Gamma (n+2) \Gamma (n+2 \varepsilon +2)s^n}{(n+2) (\varepsilon +n+2) \Gamma (n-\varepsilon +3)
		\Gamma \left(n+\varepsilon +\frac{5}{2}\right)}
\\
-\sum\limits_{n}\frac{\pi^{-1/2} (3 \varepsilon +5 n+10) \Gamma
		\left(\varepsilon +\frac{3}{2}\right)  \Gamma (n+\varepsilon +2) \Gamma (n+2 \varepsilon +2)s^n}{4 (2
		\varepsilon +1) (n+2)^2 \Gamma (2 \varepsilon ) (\varepsilon +n+2) \Gamma (n-\varepsilon +3) \Gamma \left(n+\varepsilon
		+\frac{5}{2}\right)}\, ,
\label{eqns:J10ExactSol}
\end{multline}
where we have introduced notation for multiple triangular sums
\begin{equation}
\label{TSDef}
\sum\limits_{n_1,...,n_k}f_1(n_1)...f_k(n_k) = \sum\limits_{n_k = 0}^{\infty}\sum\limits_{n_{k-1} = 0}^{n_k}...\sum\limits_{n_1 = 0}^{n_2}f_1(n_1)...f_k(n_k).
\end{equation}
Finally, the expression for $J_{11}$ master integral is obtained from the relation\footnote{It is a simple sequence of Eq. \eqref{eqns:diffeqn-s}}
\begin{equation}
J_{11} = \frac{4+s}{2(1+2\varepsilon)}\left(\varepsilon J_{10} + \frac{s}{2}\frac{d J_{10}}{ds}\right) = \frac{s(4+s)s^{-2\varepsilon}}{4(1+2\varepsilon)} \frac{d (s^{2\varepsilon}J_{10})}{ds}\, ,
\end{equation}
such that 
\begin{multline}
\frac{J_{11}\sin (\pi  \varepsilon )}{s^2\pi^{3/2} \varepsilon^3(4+s)} = \\ 
\sum\limits_{j,n}\frac{ (-2 \varepsilon +3 j+3) (-1)^{n-\varepsilon }4^{-\varepsilon +j-n-3} \Gamma \left(j-\varepsilon
		+\frac{3}{2}\right)  \Gamma (n-\varepsilon +2) \Gamma (n+\varepsilon +3) s^{n-\varepsilon }}{(j+1) (-\varepsilon
		+j+1) \Gamma \left(n+\frac{5}{2}\right) \Gamma (j-2 \varepsilon +2) \Gamma \left(n-\varepsilon
		+\frac{5}{2}\right)}
\\
+\sum\limits_{j,n}\frac{ (-1)^{n+1}4^{-\varepsilon +j-n-2}  (\varepsilon +3 j+6) \Gamma
		\left(j+\frac{5}{2}\right)  \Gamma (n+2)  \Gamma (n+2 \varepsilon +3)s^n}{(j+2) (\varepsilon +j+2) \Gamma
		\left(n+\frac{5}{2}\right) \Gamma (j-\varepsilon +3) \Gamma \left(n+\varepsilon
		+\frac{5}{2}\right)}
\\
+\sum\limits_{j,n}\frac{  (3 \varepsilon +5 j+10) (-1)^{-j+n}4^{-\varepsilon +j-n-2}
		\Gamma \left(j+\frac{5}{2}\right)  \Gamma (n+2) \Gamma (j+\varepsilon +2)  \Gamma (n+2 \varepsilon +3)s^n}{(j+2)
		\Gamma (\varepsilon ) (\varepsilon +j+2) \Gamma (j+3) \Gamma \left(n+\frac{5}{2}\right) \Gamma (j-\varepsilon +3) \Gamma
		\left(n+\varepsilon +\frac{5}{2}\right)}
\\
-\sum\limits_{n}\frac{3 \pi ^{1/2} (\varepsilon +1) (-1)^n 
		2^{-2 \varepsilon -2 n-7}  \Gamma (n+2) \Gamma (n+2 \varepsilon +3)s^n}{\Gamma (2-\varepsilon ) \Gamma \left(n+\frac{5}{2}\right) \Gamma
		\left(n+\varepsilon +\frac{5}{2}\right)}
\\
+\sum\limits_{n}\frac{ 2^{-2 \varepsilon -4} (-1)^n  (\varepsilon +3 n+6)  \Gamma (n+2) \Gamma (n+2 \varepsilon +3)s^n}{(n+2) (\varepsilon +n+2) \Gamma (n-\varepsilon +3)
		\Gamma \left(n+\varepsilon +\frac{5}{2}\right)}
\\
-\sum\limits_{n}\frac{ 2^{-2 \varepsilon -4}  (3 \varepsilon +5 n+10) \Gamma (n+\varepsilon +2) \Gamma (n+2 \varepsilon +3)s^n }{(n+2)^2 \Gamma (\varepsilon )
		(\varepsilon +n+2) \Gamma (n-\varepsilon +3) \Gamma \left(n+\varepsilon +\frac{5}{2}\right)}\, .\\
\label{eqns:J11ExactSol}
\end{multline}
Note, that the obtained solutions \eqref{eqns:J10ExactSol} and \eqref{eqns:J11ExactSol} are also valid only in the region $|s| \leq 1$.  Still, within this region the results obtained both for non-elliptic and elliptic master integrals together with their $\varepsilon$ expansion can be calculated with a very high precision with the help of \texttt{SummerTime} package \cite{SummerTime}. The latter property is due to the fact that the all arising sums are triangle with a factorized summand. The $\varepsilon$ expansion of the triangle sums is obtained by a corresponding expansion of their summands. If necessary, one can extract several terms out of the sum sign in order to explicitly separate the divergences in $\varepsilon$. Numerically, the series in $\varepsilon$ can be obtained from the exact solution with the help of the function \texttt{TriangleSumSeries} from the \texttt{SummerTime} package \cite{SummerTime}. In the latter case, the expansions up to $\varepsilon^3$ order with the accuracy of thousand digits for coefficients can be obtained in less than 10 seconds on an average computer. Similar to the case of non-elliptic canonical master integrals all 1-dimensional sums can be further rewritten in terms of generalized hypergeometric functions. To rewrite 2-dimensional sums generalized hypergeometric functions are not sufficient and we need to introduce their generalization to two variables. The latter are known as generalized Kampé de Fériet functions and are defined as \cite{MR834385}
\begin{equation}
\Fkdf{p:q:k}{l:m:n}{(a_p)}{(\alpha_l)}{(b_q)}{(\beta_m)}{(c_k)}{(\gamma_n)}{x ; y} = \sum\limits_{r,s = 0}^{\infty}\frac{\prod\limits_{j=1}^p(a_j)_{r+s}\prod\limits_{j=1}^q(b_j)_{r}\prod\limits_{j=1}^k(c_j)_{s}}{\prod\limits_{j=1}^l(\alpha_j)_{r+s}\prod\limits_{j=1}^m(\beta_j)_{r}\prod\limits_{j=1}^n(\gamma_j)_{s}}\frac{x^r}{r!}\frac{y^s}{s!}.
\end{equation}
This way for our elliptic master integrals we get
\begin{align}
\frac{J_{10}}{s^2 \varepsilon^3(1+2\varepsilon)} =  &
-\frac{\pi ^{3/2}  (\varepsilon+6) \text{csc} (\pi  \varepsilon ) \Gamma (2 \varepsilon+2)}{2^{2 \varepsilon +3}(\varepsilon +2) \Gamma (3-\varepsilon )
\Gamma \left(\varepsilon +\frac{5}{2}\right)}\times \nonumber\\
&\times\Bigg\{ \Fkdf{2:1:5}{2:0:4}{2, 2(1+\ep)}{\frac{5}{2},\frac{5}{2}+\ep}{1}{-}{1,2,\frac{5}{2},2+\ep,3+\frac{\ep}{3}}{3,3-\ep,3+\ep,2-\frac{\ep}{3}}{-\frac{s}{4} ; -s}\nonumber\\
&-\frac{\varepsilon  (\varepsilon +1) (3 \varepsilon+10)}{2 (\varepsilon +6)} \Fkdf{2:1:6}{2:0:5}{2, 2(1+\ep)}{\frac{5}{2},\frac{5}{2}+\ep}{1}{-}{1,2,\frac{5}{2},2+\ep,2+\ep,3+\frac{3\ep}{5}}{3,3-\ep,3+\ep,3,2+\frac{3\ep}{5}}{-\frac{s}{4} ; -s}\Bigg\}\nonumber\\
&+\frac{\pi \cot (\pi  \varepsilon ) \Gamma (2 \varepsilon+3) }{3  (-s)^{\varepsilon } 4^{\varepsilon +1} (1-4 \varepsilon ^2)}
\Fkdf{2:1:5}{2:0:4}{2+\ep, 2-\ep}{\frac{5}{2},\frac{5}{2}-\ep}{1}{-}{1,1,1-\ep,\frac{3}{2}-\ep,2-\frac{2\ep}{3}}{2,2-2\ep,2-\ep,1-\frac{2\ep}{3}}{-\frac{s}{4} ; -s}\nonumber\\
&+\frac{\pi ^{3/2} 2^{-2 \varepsilon -3} (\varepsilon+6) \text{csc} (\pi  \varepsilon ) \Gamma (2 \varepsilon+2)}{(\varepsilon +2) \Gamma (3-\varepsilon )\Gamma \left(\varepsilon +\frac{5}{2}\right)}\pFq{6}{5}{1,2,2,2(1+\ep),2+\ep,3+\frac{\ep}{3}}{3,3-\ep,3+\ep,\frac{5}{2}+\ep,2+\frac{\ep}{3}}{-s}
\nonumber\\
&-\frac{\pi  \varepsilon  (3 \varepsilon +10) \text{csc} (\pi\varepsilon ) \Gamma (\varepsilon +2)}{4 \left(2\varepsilon ^2+7 \varepsilon +6\right) \Gamma(3-\varepsilon )}\pFq{7}{6}{1,2,2,2(1+\ep),2+\ep,2+\ep,3+\frac{3\ep}{5}}{3,3,3-\ep,3+\ep,\frac{5}{2}+\ep,2+\frac{3\ep}{5}}{s}
\nonumber\\
&-\frac{\pi ^{3/2} 2^{-2 (\varepsilon +2)} \text{csc} (\pi \varepsilon ) \Gamma (2 \varepsilon +3)}{\Gamma(2-\varepsilon ) \Gamma \left(\varepsilon+\frac{5}{2}\right)}\pFq{3}{2}{1,2,2(1+\ep)}{\frac{5}{2},\frac{5}{2}+\ep}{-\frac{s}{4}}
\end{align}
and
\begin{align}
\frac{J_{11}\sin (\pi  \varepsilon )}{s^2\pi^{3/2} \varepsilon^3(4+s)} = & \frac{2^{-2 (\varepsilon +3)} \varepsilon 
	(\varepsilon +1) (3 \varepsilon +10) \Gamma (2
	\varepsilon +3)}{(\varepsilon +2) \Gamma
	(3-\varepsilon ) \Gamma \left(\varepsilon
	+\frac{5}{2}\right)}\times
\nonumber\\
&\times\Bigg\{\Fkdf{2:1:6}{2:0:5}{2, 3+2\ep}{\frac{5}{2},\frac{5}{2}+\ep}{1}{-}{1,2,\frac{5}{2},2+\ep,2+\ep,3+\frac{3\ep}{5}}{3,3,3-\ep,3+\ep,2+\frac{3\ep}{5}}{-\frac{s}{4} ; -s}
\nonumber \\ 
&-\frac{2 (\varepsilon +6)}{\varepsilon  (\varepsilon
	+1) (3 \varepsilon +10)}\Fkdf{2:1:5}{2:0:4}{2, 3+2\ep}{\frac{5}{2},\frac{5}{2}+\ep}{1}{-}{1,2,\frac{5}{2},2+\ep,3+\frac{\ep}{3}}{3,3-\ep,3+\ep,2+\frac{\ep}{3}}{-\frac{s}{4} ; -s}\Bigg\}
\nonumber\\
&+\frac{(-s)^{-\varepsilon } \Gamma (4-2 \varepsilon )\Gamma (\varepsilon +3)}{96\Gamma (3-2 \varepsilon ) \Gamma\left(\frac{5}{2}-\varepsilon \right)}\Fkdf{2:1:5}{2:0:4}{2-\ep, 3+\ep}{\frac{5}{2},\frac{5}{2}-\ep}{1}{-}{1,1,1-\ep,\frac{3}{2}-\ep,2-\frac{2\ep}{3}}{2,2-\ep,2-\ep,1-\frac{2\ep}{3}}{-\frac{s}{4} ; -s}
\nonumber\\
&-\frac{2^{-2 (\varepsilon +3)} \varepsilon 
	(\varepsilon +1) (3 \varepsilon +10) \Gamma (2
	\varepsilon +3)}{(\varepsilon +2) \Gamma
	(3-\varepsilon ) \Gamma \left(\varepsilon
	+\frac{5}{2}\right)}\pFq{7}{6}{1,2,2,2+\ep,2+\ep,3+2\ep,3+\frac{3\ep}{5}}{3,3,3-\ep,3+\ep,\frac{5}{2}+\ep,2+\frac{3\ep}{5}}{s}
\nonumber\\
&+\frac{2^{-2 \varepsilon -5} (\varepsilon +6) \Gamma (2
	\varepsilon +3)}{(\varepsilon +2) \Gamma
	(3-\varepsilon ) \Gamma \left(\varepsilon
	+\frac{5}{2}\right)}\pFq{6}{5}{1,2,2,2+\ep,3+2\ep,3+\frac{\ep}{3}}{3,3-\ep,3+\ep,\frac{5}{2}+\ep,2+\frac{\ep}{3}}{-s}
\nonumber\\
&-\frac{2^{-2 \varepsilon -5} (\varepsilon +1) \Gamma
	(2 \varepsilon +3)}{\Gamma (2-\varepsilon ) \Gamma
	\left(\varepsilon +\frac{5}{2}\right)}\pFq{3}{2}{1,2,3+2\ep}{\frac{5}{2},\frac{5}{2}+\ep}{-\frac{s}{4}}.
\end{align}

\section{Conclusion and discussion}\label{conclusion&Discussion}

Let us now discuss the obtained results in the context of strong and weak sides of each method used. As criteria we will consider the complexity of obtaining the results, the convenience of numerical computation and the applicability of the method to other cases. We want to emphasize that the presented discussion in first place is made for the specific example of a non-planar elliptic vertex considered in this paper. For simpler non-elliptic Feynman integrals or more complex integrals with multiple elliptic structures the features of each method may be different. It is also possible that some of methods may not be applicable at all. Nevertheless, we believe that the discussions given below can also be relevant for a wide class of elliptic Feynman integrals.

As for the difficulty in getting results the most difficult is the direct integration over Feynman parameters. The difficulty in this case is related to the reduction of individual MPLs to canonical form before next integration can performed. Moreover, this complexity grows factorially with the increase of MPLs weight lengths. Calculating the $\varepsilon^1$ corrections for considered non-planar vertex master integrals is already very difficult. The next in complexity is the evaluation of master integrals with differential equation method  in terms of iterated integrals with algebraic kernels. After we have reduced the differential system to $\varepsilon$-form, its subsequent integration goes straightforwardly. Obtaining higher  $\varepsilon$-corrections is not as difficult as in the case of direct integration. The most "simple" in the sense of getting  results is the exact Frobenius method. The solution of the first-order linear difference equations does not cause any difficulties. Moreover, the resulting solution is exact and immediately contains all $\varepsilon$-corrections. 

Now consider numerics.  The results in the form of iterated integrals with algebraic kernels can be conveniently calculated with use of existing numerical libraries for MPLs and standard integration techniques for the last "closing" integration. The results obtained in terms of triangular sums with factorized summand can be numerically evaluated\footnote{Using known acceleration techniques for such sums.} with a very high precision of several thousand digits \cite{SummerTime} inside the interval $|s| \le 1$. To compute these sums outside this interval we need to solve the problem of their analytical continuation. One way to do it is to rewrite the obtained sums in terms of well known special functions. The 1-dimensional sums can be easily written in terms of generalized hypergeometric  $_p F_q$ functions, while for 2-dimensional sums we were required to introduce their generalization to two variables - Kampé de Fériet functions. At the same time,  the results in terms of iterated integrals with algebraic kernels can be used in the whole $s$ range. However, using available techniques in the latter case it is problematic to obtain the results with arbitrary prescribed precision. In the future, we plan to study the presented functions in more detail and develop new methods for their numerical calculation.   

As for the applicability of the presented methods to other cases the choice of the method will greatly depend on the problem under consideration. For problems with single dimensionless variable the best choice seems to be a method of differential equations. It is both simpler and more universal. On the other hand for problems with multiple kinematical invariants the direct integration may be advantageous in some cases. For already studied elliptic master integrals the differential equation method works in all situations where direct integration works. However, there are cases such as considered in \cite{kites-elliptic}, where we were unable to successfully apply direct integration. Also, the solution of differential equations in terms of iterated integrals is more convenient in situation when considered master integrals occur as subgraphs in more complicated master integrals. The analytical solution in terms of power series is less studied of all three methods. In the considered problem the difference equations for power coefficients were of first order. They will be of first order also for elliptic master integrals considered in \cite{KKOVelliptic1,KKOVelliptic2}, but of course in general  difference equations of higher order should appear. Still, we think this direction deserves further study and we suppose to return to this problem in future. It is also interesting to consider solution of dimensional recurrence relations \cite{dimrecurrence} for the considered non-planar elliptic vertex along the lines of \cite{dimrecSunrise}.     

Finally, let us discuss the compactness of the obtained results. For the first sight the result of direct integration is most compact, followed by the result obtained with differential equations and at the end comes Frobenius series solution. However, if we count the number of functions in the answer, then the situation turns out to be exactly the opposite. The smallest number of functions is in the exact series solution\footnote{Here by function we mean single triangle sum.}, next comes the result of differential equation method and finally goes the result of direct integration. If we consider the higher $\varepsilon$-corrections, the situation becomes even more dramatic. The $\varepsilon^1$ corrections obtained by direct integration or differential equation methods are already too large to be presented in the paper.  On the contrary, the results obtained with the Frobenius method are exact and therefore already contains all $\varepsilon$-corrections in a rather compact form. We have also checked, that the results obtained with three different methods agree numerically with each other and with the results of sector decomposition method as implemented in FIESTA package \cite{Fiesta4}

Thus, all three methods have their specific advantages and disadvantages. And when choosing a technique, one needs to be guided by the needs of a specific calculation. Also, it is not necessary to stick with one specific method. It is quite possible to use different methods and combine them with each other.

The authors are gratefull to A.V.Kotikov, R.N.Lee and O.L. Veretin for interesting and stimulating discussions. This work was supported in part by the Foundation for the
Advancement of Theoretical Physics and Mathematics "BASIS" and Russian Science Foundation, grant 20-12-00205. The authors also would like to thank Heisenberg-Landau program.

\appendix

\section{Notation for iterated integrals}\label{notation}

Let us review here the notation used to write down the results in terms of iterated integrals with algebraic kernels. The latter are written in the following form: 
\begin{equation}
J(\underbrace{\overbrace{\Omega}^{1-form~ in~ y}}_{or~ 1-form~ in ~t},\overbrace{\omega_1^s,\ldots ,\omega_n^s}^{1-forms~ in~ s},\underbrace{\omega_1^z,\ldots ,\omega_m^z}_{1-forms~ in~ z},\overbrace{\omega_1^y,\ldots ,\omega_l^y}^{1-forms~ in~ y};s)
\end{equation}
where $\omega_i^s$, $\omega_i^z$ and , $\omega_i^y$ are some 1-forms in $s$, $z$ and $y$ correspondingly and 1-form $\Omega$ is either a 1-form with respect to the variable $y$ or a 1-form with respect to variable $t$. 1-forms $\omega_i^s$, $\omega_i^z$ and , $\omega_i^y$ form iterated integrals, and the last 1-form $\Omega$ is "closing" - the integration over it goes from 2 to infinity in the case of variable y and from 0 to 1 in the case of variable $t$.
It should be noted that the variable $z$ is not independent and depends on variables $s$ and $y$ as  
\begin{equation}
z=\frac{1}{2}\left(1+ \sqrt{1+\frac{y^2}{s}}\right)
\end{equation}
Using this relation the differential forms $\omega_i^s$ and $\omega_i^z$ can be rewritten in terms of each other. The 1-form "$\Omega$" with respect to $y$ in our expressions is of the form  ($J_y=\frac{4}{y^2\sqrt{y^2-4}}$):
\begin{align}
M_m^n = \frac{(1-z)^m z^m J_y dy}{y^2(2z-1)^n},
\end{align}
while  1-forms "$\Omega$" with respect to $t$ $(\bar{t} =1-t)$ are defined as
\begin{align}
N_{a,b}^{g} &=-\frac{g t^a\bar{t}^bdt}{1+st\bar{t}}, & F_{a,b}^{g} &=\frac{g t^a\bar{t}^bdt}{(1+st\bar{t})^2}, &\Omega_{a,b}^{g} &=\frac{-g t^a\bar{t}^bdt}{(1+st)(1+st\bar{t})}, & K_{a,b}^{g} &=\frac{g t^a\bar{t}^bdt}{(1+s\bar{t})(1+st\bar{t})}, &
\end{align}
Here, $g$ are some algebraic functions of $s$ and $t$. In our expressions $g$ are given by monomials in variables $\{1,z_1,z_2,z_3,z_4,z_5 \}$, where
\begin{equation}
\label{roots}
z_1=\sqrt{\frac{s-1/\bar{t}}{s}}, ~ z_2=\sqrt{\frac{s-1/t}{s}}, ~ z_3=\sqrt{\frac{s+1/t}{s}}, ~ z_4=\sqrt{\frac{s+1/\bar{t}}{s}}, ~ z_5=\sqrt{\frac{s+1/\bar{t}t}{s}}.
\end{equation}
Finally, 1-forms "$\omega$" with respect to $s$, $z$ or $y$ are defined as
\begin{align}
\omega_c^s &= \frac{ds}{s-c}, & \omega_c^{z_i} &=\frac{z_i ds}{s-c}, & \omega_c^z &=\frac{dz}{z-c},  & \omega_c^y &=\frac{dy}{y-c}
\end{align}
For example with the introduced notation we have
\begin{equation}
J\left(M_2^1,\omega _a^z,\omega _b^z,\omega _c^z;s\right) = \int_2^{\infty}\frac{(1-z)^2z^2J_{y}dy}{y^2(2z-1)}\int_0^z\frac{dz'}{(z'-a)}\int_0^{z'}\frac{dz''}{z''-b}\int_0^{z''}\frac{dz'''}{z'''-c}
\end{equation} 
\begin{equation}
J\left(M_1^2,\omega _a^z,\omega _b^y,\omega _c^y;s\right) = \int_2^{\infty}\frac{(1-z)zJ_{y}dy}{y^2(2z-1)^2}\int_0^z\frac{dz'}{(z'-a)}\int_0^{y}\frac{dy'}{y'-b}\int_0^{y'}\frac{dy''}{y''-c}
\end{equation} 
\begin{equation}
J\left(F^{z_5}_{-1,0},\omega _a^{z_3},\omega _b^{z_2},\omega _c^s;s\right) = \int_0^{1}\frac{\sqrt{\frac{s+1/\bar{t}t}{s}} dt}{t(1+st\bar{t})}\int_0^s\frac{\sqrt{\frac{s'+1/t}{s'}}ds'}{(s'-a)}\int_0^{s'}\frac{\sqrt{\frac{s''-1/t}{s''}}ds''}{s''-b}\int_0^{s''}\frac{ds'''}{s'''-c}
\end{equation} 
\begin{equation}
J\left(\omega _a^{z_1},\omega _b^{z_4},\omega _c^s;s\right) = \int_0^s\frac{\sqrt{\frac{s'-1/\bar{t}}{s'}}ds'}{(s'-a)}\int_0^{s'}\frac{\sqrt{\frac{s''+1/\bar{t}}{s''}}ds''}{s''-b}\int_0^{s''}\frac{ds'''}{s'''-c}
\end{equation}

\section{Effective master integrals}\label{effective-masters}

The expressions for effective master integrals for integral family \eqref{eqns:effFamily} are given by
\begin{equation}
I^{\rm eff}_{1,0,1,0,0}(s,t)=\frac{1}{\varepsilon ^2}+\frac{2}{\varepsilon }+\frac{\pi ^2}{6}+3 + \mathcal{O}(\varepsilon)
\end{equation}

\begin{multline}
I^{\rm eff}_{0,0,1,1,1}(s,t) = -\frac{1}{\varepsilon ^2}+\frac{J\left(\omega _0;s\right)-i \pi -3+\log (4)}{\varepsilon
}+\frac{1}{2} \Big(2 i \pi  J\left(\omega _0;s\right)+6 J\left(\omega _0;s\right)-\\-2
J\left(\omega _0,\omega _0;s\right)-\log (16) J\left(\omega _0;s\right)+\pi ^2-6 i
\pi -14-4 \log ^2(2)+\\+i \pi  \log (16)+12 \log
(2)\Big) + \mathcal{O}(\varepsilon)
\end{multline}

\begin{multline}
I^{\rm eff}_{1,0,1,1,0}(s,t) = -\frac{1}{\varepsilon ^2}+\frac{-st+s-3}{\varepsilon }+\frac{13 s
	\bar{t}}{2}-\frac{\pi ^2}{6}-6+z_1 (-2 s \bar{t}-1) J\left(\omega
_{1/\bar{t}}^{z_1};s\right)+
\\
+\left(2-\frac{1}{2 s \bar{t}}\right) J\left(\omega _{1/\bar{t}}^{z_1},\omega
_{1/\bar{t}}^{z_1};s\right) + \mathcal{O} (\varepsilon)
\end{multline}

\begin{multline}
I^{\rm eff}_{2,0,1,1,0}(s,t) = \frac{1}{2 \varepsilon ^2}+\frac{1}{2 \varepsilon
}+\left(\frac{1}{2 s \bar{t}}-1\right) J\left(\omega _{1/\bar{t}}^{z_1},\omega
_{1/\bar{t}}^{z_1};s\right)+\\+z_1 J\left(\omega
_{1/\bar{t}}^{z_1};s\right)+\frac{1}{12} \left(\pi ^2-6\right) + \mathcal{O}(\varepsilon)
\end{multline}

\begin{multline}
I^{\rm eff}_{1,1,0,0,1}(s,t) = -\frac{1}{\varepsilon ^2}+\frac{s t-3}{\varepsilon }+\left(2-\frac{1}{2 s t}\right) J\left(\omega _{1/t}^{z_2},\omega
_{1/t}^{z_2};s\right)-z_2 (2 s t+1) J\left(\omega
_{1/t}^{z_2};s\right)\\ +\frac{13 s
	t}{2}-\frac{\pi ^2}{6}-6 + \mathcal{O}(\varepsilon)
\end{multline}

\begin{multline}
I^{\rm eff}_{2,1,0,0,1}(s,t)=\frac{1}{2 \varepsilon ^2}+\frac{1}{2 \varepsilon
}+\left(\frac{1}{2 s t}-1\right) J\left(\omega _{1/t}^{z_2},\omega
_{1/t}^{z_2};s\right)+\\+z_2 J\left(\omega
_{1/t}^{z_2};s\right)+\frac{1}{12} \left(\pi ^2-6\right) + \mathcal{O}(\varepsilon)
\end{multline}

\begin{multline}
I^{\rm eff}_{1,0,1,1,1}(s,t)=\frac{1}{2 \varepsilon ^2}+\frac{-J\left(\omega _0;s\right)+i \pi
	+\frac{5}{2}-\log (4)}{\varepsilon }-\frac{\bar{t} J\left(\omega _{-1/t}^{z_3},\omega _{-1/t}^{z_1
		z_3},\omega _{1/\bar{t}}^{z_1};s\right)}{2 s t}\\ -\frac{J\left(\omega
	_{-1/t}^{z_3},\omega _{1/\bar{t}}^{z_1 z_3},\omega
	_{1/\bar{t}}^{z_1};s\right)}{2 s}-
-\bar{t} z_3 J\left(\omega
_{-1/t}^{z_1 z_3},\omega _{1/\bar{t}}^{z_1};s\right)-t z_3
J\left(\omega _{1/\bar{t}}^{z_1 z_3},\omega
_{1/\bar{t}}^{z_1};s\right)\\ +\frac{J\left(\omega
	_{-1/t}^{z_3},\omega _{-1/t}^{z_3},\omega _0;s\right)}{2 s t}+z_3
J\left(\omega _{-1/t}^{z_3},\omega _0;s\right)
+z_3 (\log (4)-i \pi )
J\left(\omega _{-1/t}^{z_3};s\right)\\ +\frac{(\log (4)-i \pi ) J\left(\omega
	_{-1/t}^{z_3},\omega _{-1/t}^{z_3};s\right)}{2 s t}  +J\left(\omega
_{1/\bar{t}}^{z_1},\omega
_{1/\bar{t}}^{z_1};s\right)+J\left(\omega _0,\omega _0;s\right) \\ 
-(5+i \pi
-\log (4)) J\left(\omega _0;s\right) -\frac{7 \pi
	^2}{12}+\frac{19}{2}+2 \log ^2(2)-\\-10\log (2)- i \pi  (2\log (2)-5) + \mathcal{O}(\varepsilon)
\end{multline}

\begin{multline}
I^{\rm eff}_{1,0,1,1,1}(s,t) = \frac{1}{4st}\Big[
\bar{t} J\left(\omega _{-1/t}^{z_3},\omega
_{-1/t}^{z_1 z_3},\omega _{1/\bar{t}}^{z_1};s\right)+t
J\left(\omega _{-1/t}^{z_3},\omega _{1/\bar{t}}^{z_1 z_3},\omega
_{1/\bar{t}}^{z_1};s\right)-
\\
-J\left(\omega _{-1/t}^{z_3},\omega
_{-1/t}^{z_3},\omega _0;s\right)-(\log (4)-i \pi) J\left(\omega
_{-1/t}^{z_3},\omega _{-1/t}^{z_3};s\right)\Big] + \mathcal{O}(\varepsilon)
\end{multline}

\begin{multline}
I^{\rm eff}_{1,1,0,1,1}(s,t)=\frac{1}{2 \varepsilon ^2}+\frac{-J\left(\omega
	_0;s\right)+i \pi +\frac{5}{2}-\log (4)}{\varepsilon }-\bar{t} z_4 J\left(\omega _{1/t}^{z_4 z_2},\omega
_{1/t}^{z_2};s\right)+
\\
+\frac{J\left(\omega
	_{-1/\bar{t}}^{z_4},\omega _{-1/\bar{t}}^{z_4},\omega
	_0;s\right)}{2 s\bar{t}}+J\left(\omega _{1/t}^{z_2},\omega
_{1/t}^{z_2};s\right)+z_4 J\left(\omega
_{-1/\bar{t}}^{z_4},\omega _0;s\right)+\\+z_4 (\log (4)-i \pi )
J\left(\omega _{-1/\bar{t}}^{z_4};s\right)-\frac{i (\pi +i \log (4))
	J\left(\omega _{-1/\bar{t}}^{z_4},\omega
	_{-1/\bar{t}}^{z_4};s\right)}{2 s \bar{t}}+J\left(\omega _0,\omega
_0;s\right)-
\\
-t z_4 J\left(\omega
_{-1/\bar{t}}^{z_4 z_2},\omega
_{1/t}^{z_2};s\right)-\frac{J\left(\omega
	_{-1/\bar{t}}^{z_4},\omega _{1/t}^{z_4 z_2},\omega
	_{1/t}^{z_2};s\right)}{2 s}-\frac{t J\left(\omega
	_{-1/\bar{t}}^{z_4},\omega _{-1/\bar{t}}^{z_4
		z_2},\omega _{1/t}^{z_2};s\right)}{2 s\bar{t}}+
\\
+(-5-i \pi +\log (4)) J\left(\omega _0;s\right)-\frac{7
	\pi ^2}{12}+\frac{19}{2}+\\+2 \log ^2(2)-10\log (2)- i \pi  (2\log (2)-5) + \mathcal{O}(\varepsilon)
\end{multline}

\begin{multline}
I^{\rm eff}_{2,1,0,1,1}(s,t) = \frac{1}{4s}\Big[\bar{t} J\left(\omega _{-1/\bar{t}}^{z_4},\omega
_{1/t}^{z_4 z_2},\omega _{1/t}^{z_2};s\right)+t J\left(\omega
_{-1/\bar{t}}^{z_4},\omega _{-1/\bar{t}}^{z_4
	z_2},\omega _{1/t}^{z_2};s\right)
\\
-J\left(\omega
_{-1/\bar{t}}^{z_4},\omega _{-1/\bar{t}}^{z_4},\omega
_0;s\right)-(\log (4)-i \pi ) J\left(\omega
_{-1/\bar{t}}^{z_4},\omega
_{-1/\bar{t}}^{z_4};s\right)\Big] + \mathcal{O}(\varepsilon)
\end{multline}

\begin{multline}
I^{\rm eff}_{1,1,1,0,1}(s,t) = \frac{1}{2 \varepsilon ^2}+\frac{1}{2 \varepsilon
}-J\left(\omega _{1/t}^{z_2},\omega
_{1/t}^{z_2};s\right)+\frac{J\left(\omega _0,\omega
	_{1/t}^{z_2},\omega _{1/t}^{z_2};s\right)}{2 s t}+\\+2 z_2 J\left(\omega
_{1/t}^{z_2};s\right)+\frac{\pi ^2-30}{12}  + \mathcal{O}(\varepsilon)
\end{multline} 

\begin{multline}
I^{\rm eff}_{1,1,1,1,0}(s,t) = \frac{1}{2 \varepsilon
	^2}+\frac{1}{2 \varepsilon }+\frac{J\left(\omega _0,\omega _{1/\bar{t}}^{z_1},\omega
	_{1/\bar{t}}^{z_1};s\right)}{2 s\bar{t}}-J\left(\omega
_{1/\bar{t}}^{z_1},\omega _{1/\bar{t}}^{z_1};s\right)+\\+2 z_1
J\left(\omega _{1/\bar{t}}^{z_1};s\right)+\frac{\pi ^2-30}{12} + \mathcal{O}(\varepsilon)
\end{multline} 

\begin{multline}
I^{\rm eff}_{2,1,1,1,1}(s,t)=-\frac{z_5}{8st\bar{t}(1+st\bar{t})}\Big[t^3 J\left(\omega _{-1/\bar{t}}^{z_4 z_5},\omega _{-1/\bar{t}}^{z_2
	z_4},\omega _{1/t}^{z_2};s\right)+t^2 \bar{t} J\left(\omega
_{-1/\bar{t}}^{z_4 z_5},\omega _{1/t}^{z_2 z_4},\omega
_{1/t}^{z_2};s\right)
\\
-t^2 J\left(\omega _{-1/\bar{t}}^{z_4
	z_5},\omega _{-1/\bar{t}}^{z_4},\omega _0;s\right)-t^2 J\left(\omega
_{-1/\bar{t}t}^{z_4 z_5},\omega _{-1/\bar{t}}^{z_2 z_4},\omega
_{1/t}^{z_2};s\right)+i t^2 (\pi +i \log (4)) J\left(\omega
_{-1/\bar{t}}^{z_4 z_5},\omega
_{-1/\bar{t}}^{z_4};s\right)
\\
+\bar{t}^3 J\left(\omega _{-1/t}^{z_3
	z_5},\omega _{-1/t}^{z_1 z_3},\omega _{1/\bar{t}}^{z_1};s\right)+t
\bar{t}^2 J\left(\omega _{-1/t}^{z_3 z_5},\omega
_{1/\bar{t}}^{z_1 z_3},\omega
_{1/\bar{t}}^{z_1};s\right)-\bar{t}^2 J\left(\omega _{-1/t}^{z_3
	z_5},\omega _{-1/t}^{z_3},\omega _0;s\right)
\\
-\bar{t}^2 J\left(\omega
_{-1/\bar{t}t}^{z_3 z_5},\omega _{-1/t}^{z_1 z_3},\omega
_{1/\bar{t}}^{z_1};s\right)+i \bar{t}^2 (\pi +i \log (4)) J\left(\omega
_{-1/t}^{z_3 z_5},\omega _{-1/t}^{z_3};s\right)-t \bar{t}
J\left(\omega _{-1/\bar{t}t}^{z_5},\omega _{1/t}^{z_2},\omega
_{1/t}^{z_2};s\right)
\\
-t \bar{t} J\left(\omega _{-1/\bar{t}t}^{z_5},\omega _{1/\bar{t}}^{z_1},\omega
_{1/\bar{t}}^{z_1};s\right)-t \bar{t} J\left(\omega _{-1/\bar{t}t}^{z_3 z_5},\omega _{1/\bar{t}}^{z_1 z_3},\omega
_{1/\bar{t}}^{z_1};s\right)-t \bar{t} J\left(\omega _{-1/\bar{t}t}^{z_4 z_5},\omega _{1/t}^{z_2 z_4},\omega
_{1/t}^{z_2};s\right)
\\
+t J\left(\omega _{-1/\bar{t}t}^{z_4
	z_5},\omega _{-1/\bar{t}}^{z_4},\omega _0;s\right)+\bar{t} J\left(\omega
_{-1/\bar{t}t}^{z_3 z_5},\omega _{-1/t}^{z_3},\omega _0;s\right)+t
(\log (4)-i \pi ) J\left(\omega _{-1/\bar{t}t}^{z_4 z_5},\omega
_{-1/\bar{t}}^{z_4};s\right)
\\
+\bar{t} (\log (4)-i \pi ) J\left(\omega
_{-1/\bar{t}t}^{z_3 z_5},\omega _{-1/t}^{z_3};s\right)\Big] + \mathcal{O}(\varepsilon)
\end{multline}
The accompanying {\it Mathematica} notebook contains also expressions for extra terms in $\varepsilon$-expansion.

\bibliographystyle{hieeetr}
\bibliography{litr}

\end{document}